\documentclass{article}

\usepackage{arxiv}

\usepackage[utf8]{inputenc} 
\usepackage[T1]{fontenc}    
\usepackage{hyperref}       
\usepackage{url}            
\usepackage{booktabs}       
\usepackage{amsfonts}       
\usepackage{nicefrac}       
\usepackage{microtype}      
\usepackage{lipsum}
\usepackage{graphicx}
\graphicspath{ {./images/} }
\usepackage{epstopdf, epsfig}
\usepackage{caption}
\usepackage{color, soul}
\usepackage[colorinlistoftodos]{todonotes}

\usepackage{longtable} 
\usepackage{multirow} 
\usepackage{amsmath} 
\usepackage[T1]{fontenc} 
\usepackage{tabularx} 
\usepackage{changepage} 
\usepackage{natbib} 
\usepackage{float} 
\usepackage[running]{lineno} 
\usepackage{xcolor} 
\usepackage{relsize} 
\usepackage{url} 
\usepackage{appendix} 
\title{\rm{Persistence analysis of velocity and temperature fluctuations in convective surface layer turbulence}}

\author{
Subharthi Chowdhuri \\
  Indian Institute of Tropical Meteorology\\
  Ministry of Earth Sciences, India\\
  \texttt{subharthi.cat@tropmet.res.in} \\
   \And
Tam\'{a}s Kalm\'{a}r-Nagy \\
  Department of Fluid Mechanics, Faculty of Mechanical Engineering\\
  Budapest University of Technology and Economics, Budapest, Hungary\\
  \texttt{physfluids@kalmarnagy.com} \\
  \And
Tirtha Banerjee \\
Department of Civil and Environmental Engineering\\
University of California, Irvine, CA 92697, USA \\ 
 \texttt{tirthab@uci.edu} \\
}

\begin{document}
\maketitle
\begin{abstract}
Persistence is defined as the probability that the local value of a fluctuating field remains at a particular state for a certain amount of time, before being switched to another state. The concept of persistence has been found to have many diverse practical applications, ranging from non-equilibrium statistical mechanics to financial dynamics to distribution of time scales in turbulent flows and many more. In this study, we carry out a detailed analysis of the statistical characteristics of the persistence probability density functions (PDFs) of velocity and temperature fluctuations in the surface layer of a convective boundary layer, using a field-experimental dataset. Our results demonstrate that for the time scales smaller than the integral scales, the persistence PDFs of turbulent velocity and temperature fluctuations display a clear power-law behaviour, associated with self-similar eddy cascading mechanism. Moreover, we also show that the effects of non-Gaussian temperature fluctuations act only at those scales which are larger than the integral scales, where the persistence PDFs deviate from the power-law and drop exponentially. Furthermore, the mean time scales of the negative temperature fluctuation events persisting longer than the integral scales are found to be approximately equal to twice the integral scale in highly convective conditions. However, with stability this mean time scale gradually decreases to almost being equal to the integral scale in the near neutral conditions. Contrarily, for the long positive temperature fluctuation events, the mean time scales remain roughly equal to the integral scales, irrespective of stability.
\end{abstract}


\section{Introduction}
\label{Intro}
Let $f(t)$ denote a stochastic signal fluctuating in time governed by a particular dynamics. The persistence is then the probability $P(t)$ that the quantity $f(t)-\overline{f(t)}$ does not change sign up to the time $t$ \citep{majumdar1999persistence,bray2013persistence}, where the overbar denotes the time average. Despite its simple description, only for some specific systems, such as those exhibiting fractional Brownian motions, the persistence probability density functions (PDFs) could be analytically shown to decay as a power-law, $P(t)\propto t^{-(1-H)}$ \citep{majumdar1999persistence,molchan1999maximum,aurzada2013persistence,aurzada2015persistence}. Here $H$ is the Hurst exponent ($0<H<1$), whose value when $0.5$ indicates simple Brownian motion. The power law form of $P(t)$ dictates that as the $H$ values get larger the persistence PDFs decrease more slowly, which seems consistent with the general notion that a stochastic signal displays anti-persistent or persistent behaviour depending on whether $0<H<1/2$ or $1/2<H<1$ \citep{wang2000hurst}. However, for other complex systems, no theoretical solutions exist for the persistence PDFs and these need to be computed empirically from the experimental data at hand \citep{perlekar2011persistence}. Notwithstanding the theoretical challenges, the concept of persistence has many practical applications, such as in the field of biology where one can ask how long does it take for an epidemic to spread \citep{sornette2006critical}, or in financial markets to assess when does a preferred stock will cross a threshold price \citep{zheng2002persistence}, or in the field of geophysics to predict when will the next earthquake have a dangerously high magnitude \citep{corral2004long}. Note that, depending on the context, the persistence could also be referred to as distributions of the first-passage time, or survival probability distributions, or return-time distributions, or the distributions of the inter-arrival times between the successive zero-crossings \citep{grebenkov2020preface}.

In turbulent flows, the interest in the concept of persistence or zero-crossings grew with the analytical result from \citet{rice1945mathematical}, through which it was possible to show that the frequency of the zero-crossings in a turbulent signal was related to the Taylor microscale \citep{sreenivasan1983zero}. This connection was intriguing, because it implied the dissipation rate of the turbulent kinetic energy could be directly computed from the zero-crossing frequencies. Since then, several studies are carried out in wall-bounded turbulent flows to verify this result and the agreements obtained with the theoretical prediction are more or less satisfactory \citep{antonia1976bursts,narayanan1977experiments,sreenivasan1983zero,poggi2010evaluation}. However, there has been a fair amount of disagreement among different experiments regarding the statistical characteristics of the PDFs of the inter-arrival times between the successive zero-crossings (hereafter, the persistence PDFs). \citet{narayanan1977experiments} found that in a turbulent boundary layer, the persistence PDFs of the velocity fluctuations were log-normal to a good approximation. Later, \citet{sreenivasan1983zero} and \citet{kailasnath1993zero} found the persistence PDFs of the velocity fluctuations and momentum flux signals were double-exponential in nature. Their interpretation of this behaviour was that the long intervals are a consequence of the large-scale structures passing the sensor and the short intervals are a consequence of the impinging small-scale motions superposed on the large-scale structures. Subsequently, \citet{bershadskii2004clusterization} showed that the persistence PDFs of temperature fluctuations from a turbulent convection experiment followed a power-law distribution, which indicates scale-free behavior. In a follow up study, \citet{sreenivasan2006clustering} commented that when the temperature behaved like an active scalar in convective turbulence, the persistence PDFs followed a power-law distribution. On the other hand, when the temperature behaved like a passive scalar in shear-driven turbulence, the persistence PDFs followed a log-normal distribution. Recently, \citet{kalmar2019complexity} noted that the persistence PDFs of velocity fluctuations followed log-normal distribution in a turbulent flow around a street canyon. 

In atmospheric turbulence the investigation of the statistical properties of the persistence PDFs of turbulent fluctuations is quite rare. Nevertheless, there are a few limited studies available from the atmospheric surface layer which report the persistence PDFs of velocity and scalar fluctuations \citep{yee1993statistical,yee1995measurements,cava2009effects,cava2012role}. The atmospheric surface layer is a generalization of the inertial layer of unstratified wall-bounded flows by including the effect of buoyancy, where the effects of surface roughness are no longer important and the modulations by the outer eddies are not too strong \citep{barenblatt1996scaling,davidson2015turbulence}. \citet{yee1993statistical,yee1995measurements} reported that the persistence PDFs of scalar concentrations displayed a double power-law in a near-neutral surface layer. Later, \citet{katul1994conditional} also observed the same, when they investigated the persistence PDFs of the burst events in the sensible heat flux in a convective surface layer. \citet{pinto2014double} have shown that the double power-law feature in a distribution function is related to the presence of two sets of fractals with two different fractal dimensions associated with two different scale free processes. However, \citet{narasimha2007turbulent} found the persistence PDFs of the momentum flux events in a near-neutral surface layer followed an exponential distribution, suggestive of a Poisson type process. Recently, \citet{cava2009effects} demonstrated that the persistence PDFs of the velocity and scalar fluctuations in a canopy surface layer turbulence could be power-laws with log-normal cut-offs or log-normal distributions depending on the measurement height in the canopy. Although in due course, \citet{cava2012role} showed that the persistence PDFs of the velocity and scalar fluctuations in canopy and atmospheric surface layer turbulence could be modelled as a power-law distribution with an exponential cut-off in convective conditions. \citet{chamecki2013persistence} has lent support for this model by investigating the persistence PDFs of velocity fluctuations above and within a cornfield canopy. 

Therefore, from this brief review, it is apparent that there is a very little consensus about the statistical characteristics of the persistence PDFs for both laboratory and atmospheric turbulent flows. Nevertheless, in a convective atmospheric surface layer, detailed understanding of the persistence properties of velocity and temperature fluctuations is important since it holds the key to explain the quadrant cycles of the heat and momentum fluxes. This is because, the switching patterns of the heat and momentum fluxes from one quadrant to the other are dependent on the zero-crossings of the component signals, as described by their persistence PDFs. Thus, we define our objectives for this study as:
\begin{enumerate}
    \item To carry out a detailed analysis to establish the statistical scaling properties of the persistence PDFs of velocity and temperature fluctuations in a convective surface layer.
    \item To empirically connect the statistical scaling properties of the persistence PDFs with the turbulent structures in a convective surface layer.
\end{enumerate}

The present study is organized in three different sections. In Section \ref{Data}, we provide the descriptions of the field-experimental dataset and methodology used in this study, in Section \ref{results} we present and discuss the results, and lastly in Section \ref{conclusion} we conclude by presenting our findings and providing the future research direction.

\section{Materials and methods}
\label{Data}
In this study, the dataset being used is from the Surface Layer Turbulence and Environmental Science Test (SLTEST) experiment. The SLTEST experiment ran continuously for nine days from 26 May 2005 to 03 June 2005, over a flat and homogeneous terrain at the Great Salt Lake desert in Utah, USA (40.14$^\circ$ N, 113.5$^\circ$ W), with the aerodynamic roughness length ($z_0$) being $z_{0}\approx$ 5 mm \citep{metzger2007near}. During this experiment, nine North-facing time synchronized CSAT3 sonic anemometers were mounted on a 30-m mast, spaced logarithmically over an 18-fold range of heights, from 1.42 m to 25.7 m, with the sampling frequency being set at 20 Hz.

During the daytime convective periods, the standard practice is to compute the turbulent statistics in the atmospheric surface layer over a 30-min period \citep{panosfsky1984atmospheric,kaimal1994atmospheric,foken2008micrometeorology}. Therefore, the data from all the nine sonic anemometers were restricted to rain-free daytime convective periods and subsequently being divided into 30-min sub-periods, containing the 20-Hz measurements of the three wind components and the sonic temperature \citep{chowdhuri2019empirical}. To select the 30-min periods for the persistence analysis, we followed the detailed data selection methods as outlined in \citet{chowdhuri2019revisiting}. Note that, we rotated the coordinate systems of all the nine sonic anemometers in the streamwise direction by applying the double-rotation method of \citet{kaimal1994atmospheric} for each 30-min period. 

A total of 261 combinations of the stability ratios ($-\zeta=z/L$, where $L$ is the Obukhov length) were possible for the selected 30-min periods from the convective conditions ($-L>0$). The stability ratio $-z/L$ is the ratio between the turbulent kinetic energy generated due to buoyancy and due to shear, with the Obukhov length ($L$) being defined as,
\begin{equation}
L = -\frac{u_*^3T_0}{k_{v}gH_0},
\label{obukhov}
\end{equation} 
where $T_{0}$ is the surface air temperature, $g$ is the acceleration due to gravity (9.8 m s$^{-2}$), $H_{0}$ is the surface kinematic heat flux, $k_{v}$ is the von K\'arm\'an constant (0.4), and $u_{*}$ is the friction velocity. It is to be noted that, these are the same set of runs used by \citet{chowdhuri2019revisiting} for their study of turbulence anisotropy. The entire range of $-\zeta$ (12 $\leq -\zeta \leq$ 0.07) was divided into six stability classes and considered for the persistence analysis (see table \ref{tab:1}). For each 30-min run, the turbulent fluctuations of the three velocity components in the streamwise ($u^{\prime}$), cross-stream ($v^{\prime}$), and vertical ($w^{\prime}$) directions, along with the fluctuations in the sonic temperature ($T^{\prime}$) were computed by removing the linear trend from the 30-min period associated with the respective variables \citep{donateo2017case}. 

\begin{table}[h]
  \caption{The six different stability classes formed from the ratio $-\zeta=z/L$ in an unstable atmospheric surface layer flow, where $z$ is the height above the surface and $L$ is the Obukhov length. The ratios span from highly convective ($-\zeta>2$) to near neutral ($0<-\zeta<0.2$) conditions. The number of 30-min runs and associated heights with each of the stability classes are given. The total number of zero-crossings (\# ZC) in $u^{\prime}$, $v^{\prime}$, $w^{\prime}$, and $T^{\prime}$ signals associated with each stability class are also provided.}
  \begin{center}
\def~{\hphantom{0}}
  \begin{tabular}{lcccccc}
   \toprule
      \multirow{2}{*}{Stability class}  & Number of & Heights & $u^{\prime}$ & $v^{\prime}$ & $w^{\prime}$ & $T^{\prime}$ \\[3pt]
      & 30-min runs & & (\# ZC) & (\# ZC) & (\# ZC) & (\# ZC)\\
      \\
      \midrule
$-\zeta \ >$ 2 & 55 & $z=$ 6.1, 8.7, 12.5, 17.9, 25.7 m & 116459 & 86576 & 142751 & 127633\\
\\
1 $< \ -\zeta \ <$ 2 & 53 & $z=$ 3, 4.3, 6.1, 8.7, 12.5, 17.9, 25.7 m & 119511 & 92236 & 189131 & 138785\\
\\
0.6 $< \ -\zeta \ <$ 1 & 41 & $z=$ 2.1, 3, 4.3, 6.1, 8.7, 12.5, 17.9 m & 97575 & 73884 & 184652 & 115173\\
\\
0.4 $< \ -\zeta \ <$ 0.6 & 34 & $z=$ 1.4, 2.1, 3, 4.3, 6.1, 8.7 m & 90551 & 70957 & 193302 & 107356\\
\\
0.2 $< \ -\zeta \ <$ 0.4 & 44 & $z=$ 1.4, 2.1, 3, 4.3, 6.1 m & 128538 & 98123 & 293816 & 154774\\
\\
0 $< \ -\zeta \ <$ 0.2 & 34 & $z=$ 1.4, 2.1, 3 m & 114383 & 95228 & 285961 & 143470\\
\bottomrule
  \end{tabular}
\label{tab:1}
  \end{center}
\end{table}

\begin{figure*}[h]
\centering
\hspace*{-0.8in}
\includegraphics[width=1.2\textwidth]{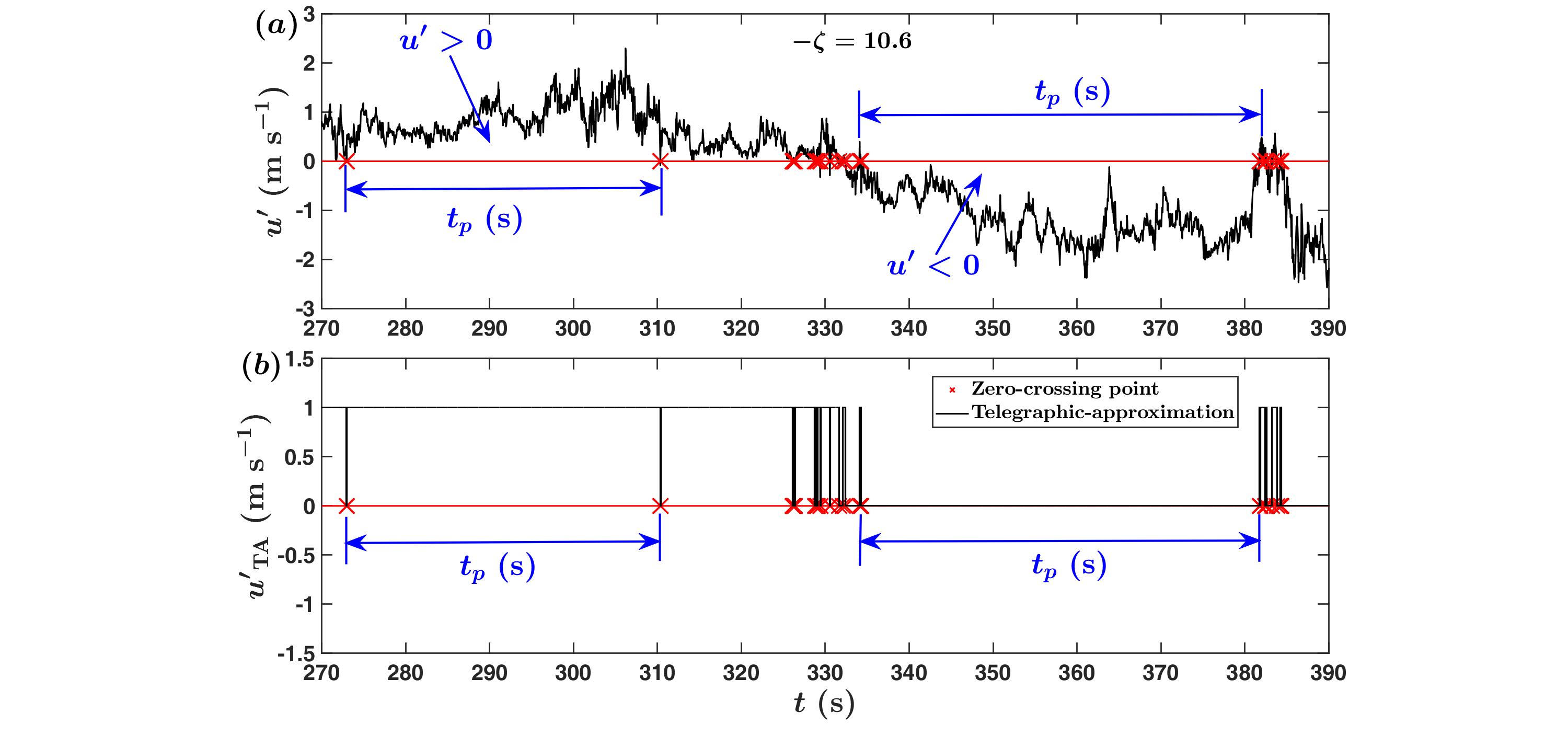}
  \caption{A 120-s long section of a time series of $u^{\prime}$ from a highly convective surface layer corresponding to a $-\zeta=$ 10.6 are shown for (a) actual values and (b) its telegraphic approximation (TA), where $u^{\prime}>0$ is denoted as 1 and $u^{\prime}<0$ is denoted as 0. The red horizontal line denotes the position of zero and the red-crosses show the points where the $u^{\prime}$ signal changes its sign from positive to negative or vice-versa (zero-crossings). To provide an example, two particular regions of $u^{\prime}$ signal are highlighted where the positive and negative values persist for a time $t_{p}$ (around 30-50 s).}
\label{fig:1}
\end{figure*}

A graphical illustration of the persistence phenomenon is provided in figure \ref{fig:1}, where a 120-s long section of $u^{\prime}$ signal is shown for a particular 30-min run, corresponding to a $-\zeta=10.6$. Figure \ref{fig:1} shows that the $u^{\prime}$ signal displays persistent positive or negative (i.e. above or below the mean) values for a particular amount of time, denoted as $t_{p}$. Note that, the persistence time $t_{p}$ can also be interpreted as the inter-arrival time between the subsequent zero-crossings where the $u^{\prime}$ signal changes its sign. The zero-crossings are identified by using the telegraphic approximation (TA) of the $u^{\prime}$ signal as,
\begin{equation}
(u^{\prime})_{\rm TA}=\frac{1}{2}(\frac{u^{\prime}(t)}{|u^{\prime}(t)|}+1),
\label{TA}
\end{equation}
and locating the points where the TA series changes its value from 0 to 1 or vice-versa (see figure \ref{fig:1}b). The associated probability that the $u^{\prime}$ signal stays positive or negative for $t_{p}$ amount of time can be evaluated by constructing the persistence PDFs using standard statistical procedures (see Appendix \ref{app_A}). Moreover, the same concept can also be extended to other turbulent fluctuations such as $v^{\prime}$, $w^{\prime}$, and $T^{\prime}$. 

For any stability class as outlined in table \ref{tab:1}, the number of persistent events (corresponding to each $t_{p}$ values) in $u^{\prime}$, $v^{\prime}$, $w^{\prime}$, and $T^{\prime}$ signals can be represented by the number of zero-crossings. In table \ref{tab:1}, we provide the number of zero-crossings or persistent events by considering all the 30-min runs from a particular stability class. Typically, we encounter in the order of 10$^{5}$ number of zero-crossings for each signal from every stability class (see table \ref{tab:1}). Therefore, the persistence PDFs of velocity and temperature fluctuations for each of these six stability classes are constructed over these large number of ensemble events, to ensure their statistical robustness. In the subsequent section, we discuss the properties of these persistence PDFs, corresponding to these six stability classes.

\section{Results and discussion}
\label{results}
Before describing the features of the persistence PDFs, it is worthwhile to discuss about the presentation of the results. Earlier studies by \citet{narayanan1977experiments}, \citet{sreenivasan1983zero}, and \citet{kailasnath1993zero} on turbulent boundary layer flows have presented the results on persistence time scales after converting those to spatial length scales by employing Taylor's frozen turbulence  hypothesis. It is important to note that although they presented results on the inter-arrival times between the two successive zero-crossings of the turbulent signals, their convention is equivalent to persistence time $t_{p}$ (see figure \ref{fig:1}). They interpreted this spatial length scale as a representative of the eddy length scale of the flow, given the connection between the mean value of the zero-crossings with the Taylor microscale \citep{antonia1976bursts,sreenivasan1983zero,kailasnath1993zero}. 

In the convective surface layer turbulence, it is a common practice to normalize the spatial length scales in the streamwise direction by the height above the surface \citep{kaimal1972spectral,kader1989spatial,kader1991spectra,mcnaughton2000power,banerjee2013logarithmic,ghannam2018scaling}. This is related to the assumption that the turbulent structures in a convective surface layer are self-smilar with height \citep{monin2013statistical,tennekes1972first,kader1991spectra,malhi2004low,mcnaughton2004attached,mcnaughton2004turbulence}. Thus, before computing the persistence PDFs, it seems apropos to convert the persistence time $t_{p}$ to a streamwise length ($t_{p}\overline{u}$, where $\overline{u}$ is the mean horizontal wind speed) by applying Taylor's frozen turbulence hypothesis and subsequently normalizing the same with $z$. Note that, for all our selected runs $\sigma_{u}/\overline{u}$ (where $\sigma_{u}$ is the standard deviation of the streamwise velocity) was less than 0.2, so the Taylor's frozen turbulence hypothesis could be assumed to be valid \citep{willis1976use}. The normalized variable $(t_{p}\overline{u})/z$ has a large range, given the minimum of $t_{p}$ is restricted by the sampling interval of 0.05 s (20-Hz sampling frequency) and the maximum of $t_{p}$ could go as large as in the order of 10$^{2}$ s (see figure \ref{fig:1}). A well-suited strategy to evaluate the PDFs of such variables is to take the logarithmic transformation and then binning the transformed variables in the logarithmic space \citep{christensen2005complexity,newman2005power,pueyo2006diversity,sims2007minimizing,benhamou2007many,white2008estimating,dorval2011estimating,newberry2019self}. More details about the effect of binning strategy on the persistence PDFs is provided in the Appendix \ref{app_A}. 

Therefore, we begin with discussing the properties of the persistence PDFs of velocity and temperature fluctuations in a convective surface layer by presenting the empirical results scaled with $z$. Subsequently, to develop a physical understanding of these persistence PDFs, we present comprehensive evidence to underpin their scaling properties and relate them with the turbulent flow structures.

\subsection{Persistence PDFs of velocity and temperature fluctuations}
\label{persistence_PDFs}
\begin{figure}[h]
\centering
\hspace*{-0.75in}
\includegraphics[width=1.25\textwidth]{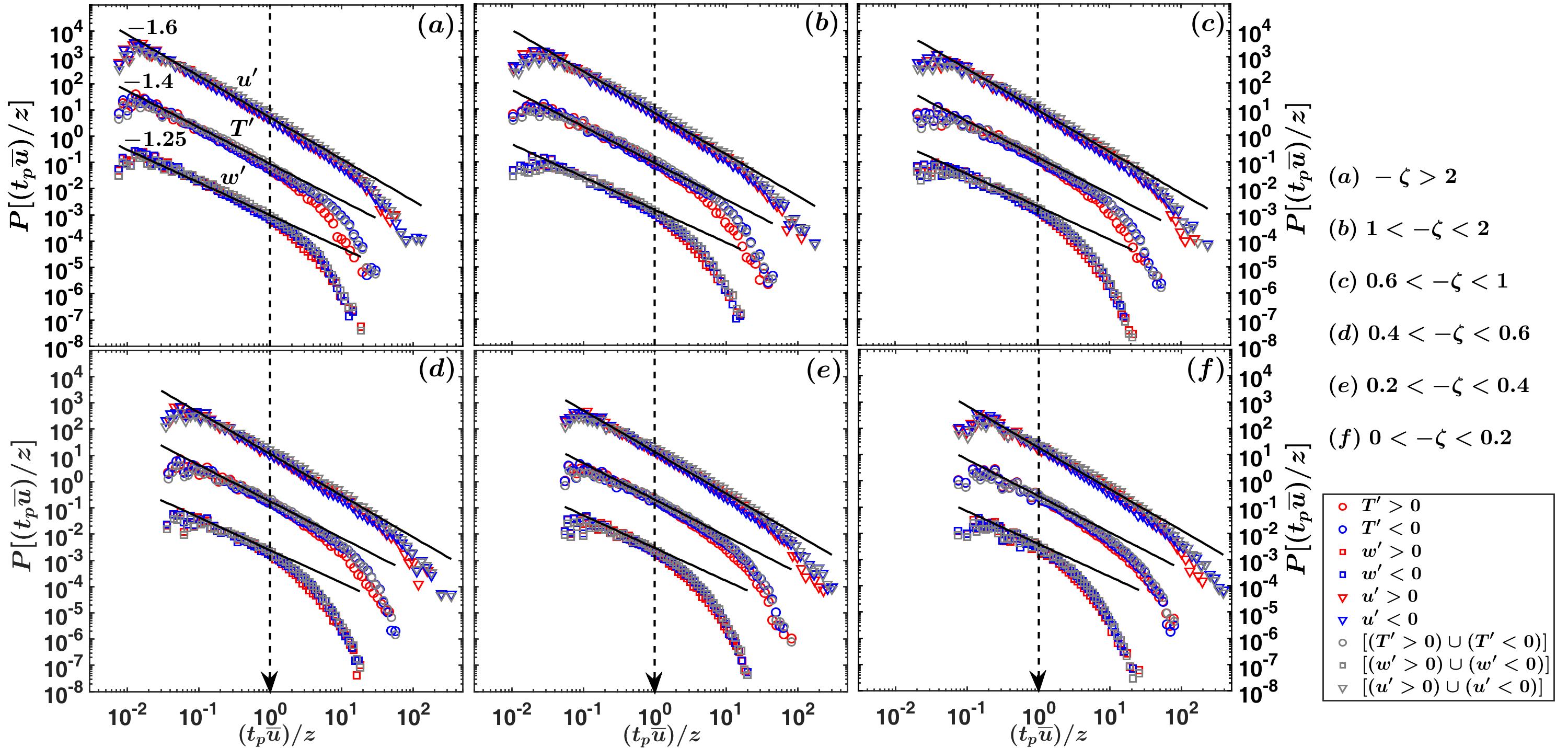}
  \caption{The persistence PDFs of the normalized streamwise sizes $(t_{p}\overline{u})/z$ corresponding to the positive (red) and negative (blue) fluctuations in the streamwise velocity ($u^{\prime}$, inverted triangles), temperature ($T^{\prime}$, circles), and vertical velocity ($w^{\prime}$, squares) are shown separately for the six different stability classes as indicated in the right most corner. The grey coloured inverted triangle, circle, and square markers on all the panels show the persistence PDFs computed after considering both the positive and negative fluctuations together. The PDFs of $w^{\prime}$, $T^{\prime}$, and $u^{\prime}$ are shifted vertically by two decades for visualization purpose and the legends corresponding to each panel are provided at the right-most bottom corner. The thick black lines on all the panels show the power-laws with their respective slopes being mentioned on the panel (a). The dashed black line denotes the value $(t_{p}\overline{u})/z=$ 1.}
\label{fig:2}
\end{figure}

In figure \ref{fig:2}, we show the persistence PDFs of velocity ($u^{\prime}$ and $w^{\prime}$) and temperature fluctuation ($T^{\prime}$) signals, plotted against $(t_{p}\overline{u})/z$ for the six different stability classes outlined in table \ref{tab:1}. These persistence PDFs ($P[(t_{p}\overline{u})/z]$) are computed after logarithmically binning the ensemble of values of $(t_{p}\overline{u})/z$ from a particular stability class and then using (\ref{pdf_conversion}) to take into account the effect of the variable bin-width owing to the logarithmic transformation (see Appendix \ref{app_A}). The log-log representation is used in figure \ref{fig:2}, such that the power-law functions on such plots would be represented by straight line segments.

Similar to \citet{bershadskii2004clusterization} and \citet{chamecki2013persistence}, the persistence PDFs are computed separately for the positive and negative fluctuations (shown as blue and red coloured markers in figure \ref{fig:2}) for $u^{\prime}$, $w^{\prime}$, and $T^{\prime}$ signals, and compared with the persistence PDFs for the total fluctuations (combining both positive and negative) shown in grey coloured markers. Note that, we focus on the $u^{\prime}$, $w^{\prime}$, and $T^{\prime}$ signals since unravelling the characteristics of their persistence PDFs have implication towards understanding the genesis of ejection and sweep cycles in the streamwise momentum ($u^{\prime}w^{\prime}$) and heat flux ($w^{\prime}T^{\prime}$) signals. Nevertheless, the persistence PDFs of the $v^{\prime}$ signals are shown in figure \ref{fig:s1} of the supplementary material and they display similar characteristic shape as the $u^{\prime}$ signals for all the six stability classes.

From figure \ref{fig:2}a, one can notice that under highly convective stratification regimes, an extensive straight line segment is observed in the persistence PDF of $u^{\prime}$ signal, with a cutoff at $(t_{p}\overline{u})/z \approx 10$. Moreover, the persistence PDFs of $T^{\prime}$ and $w^{\prime}$ signals also display a straight line segment for the small values of $(t_{p}\overline{u})/z$, although their extent remains shorter than the $u^{\prime}$ signal, with cutoffs at $(t_{p}\overline{u})/z \approx 4$ and $(t_{p}\overline{u})/z \approx 1$ respectively (figure \ref{fig:2}a). In the log-log representation, a straight line segment is reminiscent of a power-law function, thus indicating the persistence PDFs of $u^{\prime}$, $w^{\prime}$, and $T^{\prime}$ signals behave in a power-law fashion up to a certain threshold of $(t_{p}\overline{u})/z$. Note that, the exponents of these power-law functions in figure \ref{fig:2}a are different for the $u^{\prime}$, $T^{\prime}$ and $w^{\prime}$ signals, being equal to $1.6$, $1.4$, and $1.25$ respectively. These exponents have been estimated from a linear regression on the log-log plots in figure \ref{fig:7}, which we will revisit in Section \ref{scaling} while discussing the physical significance of this power-law behaviour. Since power-laws are scale invariant, a different scaling of the persistence time $t_{p}$ as employed in figure \ref{fig:8} would not affect their exponents.

To judge the effect of stability on the exponents of these power-law functions, we compared the exponents obtained from the highly convective stability (figure \ref{fig:2}a) with the other five stability classes (figures \ref{fig:2}b to \ref{fig:2}f). We note that for the $w^{\prime}$ signal, with the change in stability, the extent of this power-law region gradually shrinks, becoming almost non-existent in the near neutral stability (figures \ref{fig:2}a to \ref{fig:2}f). On the contrary, no discernible change is being observed in the power-law behaviour of the $u^{\prime}$ signals with stability. However, for the $T^{\prime}$ signals there is a modest change in the extent of the power-law region with stability, although not as clearly visible as in the $w^{\prime}$ signal. Along with that, the threshold streamwise sizes up to which the power-law behaviour holds are also not similar for these three signals. For the $u^{\prime}$ signal, at streamwise sizes greater than about $10z$, the departure from the power law is noted. Whereas for the $w^{\prime}$ signal, the threshold streamwise size is much closer to $z$. On the other hand, for the $T^{\prime}$ signal the threshold streamwise size is approximately in between of the sizes corresponding to $u^{\prime}$ and $w^{\prime}$ signals. 

Additionally, we note that for the $u^{\prime}$ and $w^{\prime}$ signals, the persistence PDFs of positive and negative fluctuations almost coincide with each other for all the values of $(t_{p}\overline{u})/z$. Nevertheless, the same is not true for the $T^{\prime}$ signal. It could be noticed that for the highly convective stability class ($-\zeta>2$, figure \ref{fig:2}a) the persistence PDFs show a slight disparity between the positive and negative $T^{\prime}$ signals at the larger values of $(t_{p}\overline{u})/z$. However, this difference gradually disappears with the change in stability from highly convective to near neutral (figures \ref{fig:2}a to \ref{fig:2}f). Therefore, one may ask \emph{for highly convective stability, what causes this discrepancy between the persistence PDFs associated with positive and negative temperature fluctuations?}  

\subsubsection{Linkage between the persistence PDFs and asymmetric distribution}
To investigate the aforementioned question, it is useful to consider a premultiplied form of the persistence PDFs such as, $(t_{p}\overline{u})/z \times P[(t_{p}\overline{u})/z]$. From phenomenological arguments, we will show that such form of the persistence PDF is equivalent to a time-fraction distribution associated with $(t_{p}\overline{u})/z$, by directly connecting the premultiplied persistence PDF with the PDF of the corresponding signal itself. We emphasize that such a connection is not possible to establish from the persistence PDFs alone (shown in figure \ref{fig:2}), without considering its premultiplied form. 

Therefore, to prove such linkage, let us consider we have an $N$-member ensemble of $(t_{p}\overline{u})/z$ values corresponding to a signal in a particular state. Note that, in the present context, there are only two possible states where the signal could either be above or below the mean (positive or negative fluctuations). Since the persistence time is computed based on the number of points lying between successive zero-crossings multiplied by the sampling interval (see figure \ref{fig:1}), we can write,
\begin{equation}
\sum\limits_{i=1}^{Q}{[(t_{p}\overline{u})/z]_{i}n_{i}} \propto T_{f},
\label{t_p}
\end{equation}
where $Q$ is the number of unique members in the $N$-member ensemble of $(t_{p}\overline{u})/z$, $n_{i}$ is the frequency of occurrences of these unique members, and $T_{f}$ is the total time-fraction spent by the signal in that particular state. The right hand side of (\ref{t_p}) stems from the fact that the left hand side of the same equation can be rearranged as,
\begin{align}
\begin{split}
\sum\limits_{i=1}^{Q}{[(t_{p}\overline{u})/z]_{i}n_{i}} &= N\sum\limits_{i=1}^{Q}{[(t_{p}\overline{u})/z]_{i}P[(t_{p}\overline{u})/z]}
\\
&= N \overline{(t_{p}\overline{u})/z},
\end{split}
\label{t_p_1}
\end{align}
where $P[(t_{p}\overline{u})/z]=(n_{i}/N)$ is the probability of selecting a unique member from the $N$-member ensemble. Given the sum of all the possible $t_{p}$ values is equal to the total time spent by the signal in a particular state and $N$, $\overline{u}$, and $z$ are constants for a particular period, the quantity $N \overline{(t_{p}\overline{u})/z}$ in (\ref{t_p_1}) is equivalent to $T_{f}$ with some proportionality constants. Thus, from (\ref{t_p}) and (\ref{t_p_1}), we can rewrite (\ref{t_p}) in the integral form as,
\begin{equation}
\int_{(\frac{t_{p}\overline{u}}{z})_{\rm min}}^{(\frac{t_{p}\overline{u}}{z})_{\rm max}} (\frac{t_{p}\overline{u}}{z}) \ P[(\frac{t_{p}\overline{u}}{z})] \ d{(\frac{t_{p}\overline{u}}{z})} \propto T_{f}.
\label{t_p1}
\end{equation}
Since (\ref{t_p1}) holds true for both positive and negative fluctuations, we may rewrite (\ref{t_p1}) as,
\begin{equation}
\int_{(\frac{t_{p}\overline{u}}{z})_{\rm min}}^{(\frac{t_{p}\overline{u}}{z})_{\rm max}} \Big( \Big [\frac{t_{p}\overline{u}}{z} P(\frac{t_{p}\overline{u}}{z}) \Big]_{\rm N}-\Big [\frac{t_{p}\overline{u}}{z} P(\frac{t_{p}\overline{u}}{z}) \Big]_{\rm P} \Big) \ d{(\frac{t_{p}\overline{u}}{z})} \propto \Delta T_{f},
\label{t_p2}
\end{equation}
where the subscripts P and N refer to the positive and negative fluctuations and $\Delta T_{f}$ is the difference in the time fractions associated with those. For a particular signal, the difference in time fractions associated with positive and negative fluctuations can be estimated from the PDF of that signal as,
\begin{equation}
\Big[\int_{-\infty}^{0} p(\hat{x}) d \hat{x}-\int_{0}^{\infty} p(\hat{x}) d\hat{x} \Big]=\Delta T_{f},
\label{t_p3}
\end{equation}
where $p(\hat{x})$ is the PDF of the signal $x$ normalized as $\hat{x}=x^{\prime}/\sigma_{x}$, such that $\hat{x}$ is a standard variable with zero mean and unit standard deviation. Through cumulant expansion of \citet{nakagawa1977prediction}, $p(\hat{x})$ can be expressed as,
\begin{align}
\begin{split}
&p(\hat{x})=G(\hat{x})\Big[1+\frac{\overline{{\hat{x}}^3}}{6}({\hat{x}}^3-3\hat{x})+\frac{1}{24}({\overline{{\hat{x}}^4}}-3)({\hat{x}}^4-6{\hat{x}}^2+3)\Big]
\\
&\quad \hspace{20mm} G(\hat{x})=\frac{1}{\sqrt{2\pi}}\exp(-\frac{{\hat{x}}^2}{2}),
\end{split}
\label{t_p4}
\end{align}
where $G(\hat{x})$ is the standard Gaussian distribution and $\overline{{\hat{x}}^3}$ and $\overline{{\hat{x}}^4}$ are the skewness and kurtosis of the signal $x$. \citet{katul1997turbulent_b} showed that if the kurtosis is ignored in (\ref{t_p4}), then the left hand side of (\ref{t_p3}) can be analytically evaluated as,
\begin{align}
\begin{split}
&\int_{-\infty}^{0} p(\hat{x}) d \hat{x}=\frac{6+(\sqrt{2/\pi})\ \overline{{\hat{x}}^3}}{12}
\\
&\quad\quad \hspace{-5mm} \int_{0}^{\infty} p(\hat{x}) d\hat{x}=\frac{6-(\sqrt{2/\pi}) \ \overline{{\hat{x}}^3}}{12}.
\end{split}
\label{t_p5}
\end{align}
By replacing the result from (\ref{t_p5}) in (\ref{t_p3}), we can write (\ref{t_p2}) as,
\begin{equation}
\int_{(\frac{t_{p}\overline{u}}{z})_{\rm min}}^{(\frac{t_{p}\overline{u}}{z})_{\rm max}} \Big( \Big [\frac{t_{p}\overline{u}}{z} P(\frac{t_{p}\overline{u}}{z}) \Big]_{\rm N}-\Big [\frac{t_{p}\overline{u}}{z} P(\frac{t_{p}\overline{u}}{z}) \Big]_{\rm P} \Big) \ d{(\frac{t_{p}\overline{u}}{z})} \propto \frac{\overline{{\hat{x}}^3}}{3\sqrt{2\pi}}.
\label{t_p6}
\end{equation}
Hence, from (\ref{t_p2})--(\ref{t_p6}) we conclude that the difference between the positive and negative fluctuations in the premultiplied form of the persistence PDFs is directly related to the non-Gaussian characteristics of the signal, approximated by its skewness. In addition to this, we also note that by performing a change of variables, \citet{dorval2011estimating} has shown that the premultiplied PDF of a stochastic variable $x$ is equivalent to the PDF of $\log(x)$, known as the logarithmic PDF (see Appendix \ref{app_A} for further details). 

\subsubsection{Premultiplied or logarithmic persistence PDFs of velocity and temperature fluctuations}
\begin{figure}[h]
\centering
\hspace*{-0.75in}
\includegraphics[width=1.25\textwidth]{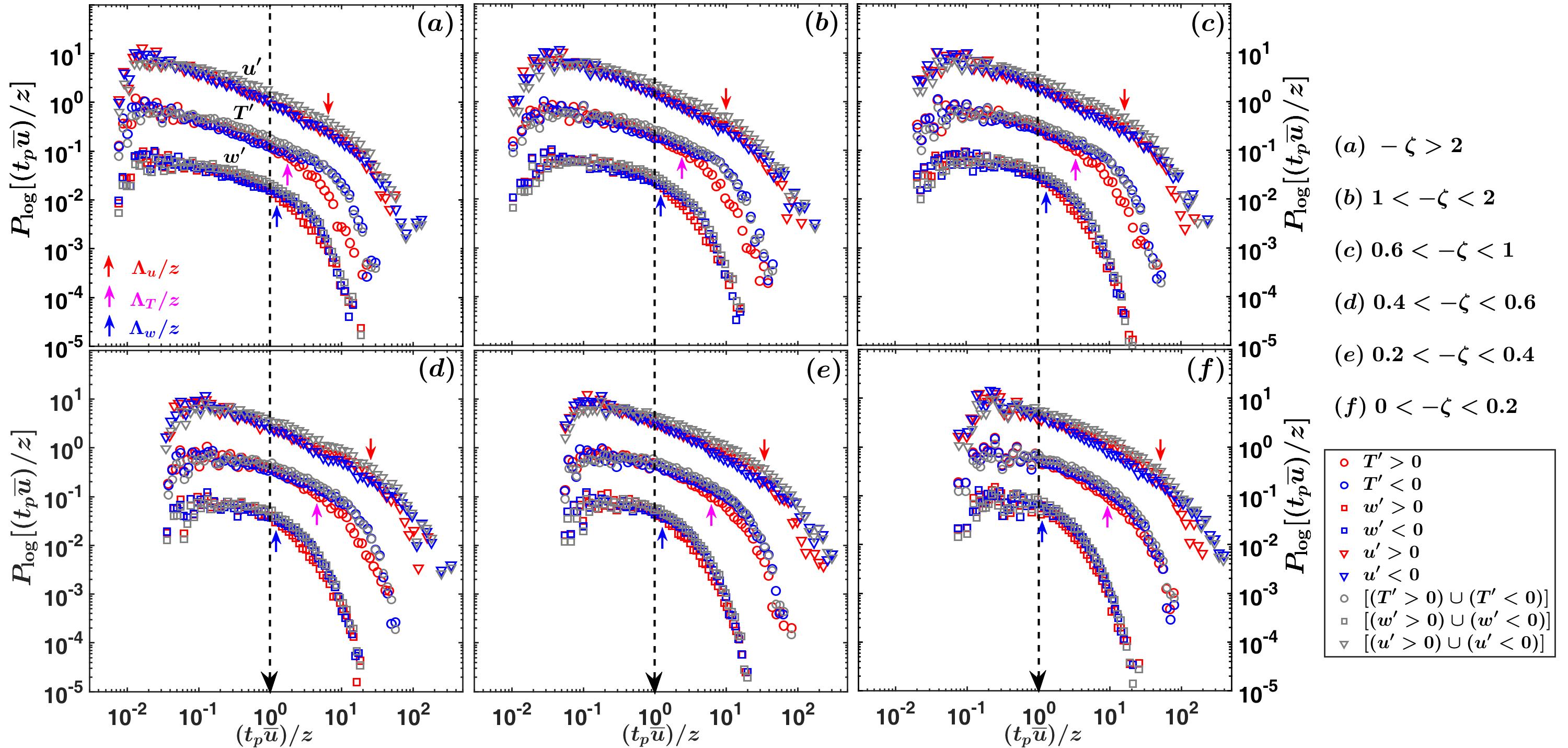}
  \caption{Same as in figure \ref{fig:2} but for the logarithmic persistence PDFs of the normalized streamwise sizes $(t_{p}\overline{u})/z$. The logarithmic PDFs of $w^{\prime}$, $T^{\prime}$, and $u^{\prime}$ are shifted vertically by a decade for visualization purpose and the legends corresponding to each panel are provided at the right-most bottom corner. The coloured arrows on all the panels show the position of the normalized integral length scales, corresponding to $w^{\prime}$ (blue arrows), $T^{\prime}$ (pink arrows), and $u^{\prime}$ (red arrows) signals.}
\label{fig:3}
\end{figure}

Figure \ref{fig:3} shows the logarithmic PDFs of $(t_{p}\overline{u})/z$ for the same six different stability classes shown in figure \ref{fig:2}. These PDFs are computed after taking the logarithm of $(t_{p}\overline{u})/z$ in the base 10 and dividing the fraction of samples by the logarithmic bin-width $d\log_{10}[(t_{p}\overline{u})/z]$. Similar to figure \ref{fig:2}, in figure \ref{fig:3} the logarithmic persistence PDFs are shown separately for the positive and negative fluctuations and compared with the total fluctuations (combining both positive and negative), corresponding to the $u^{\prime}$, $w^{\prime}$, and $T^{\prime}$ signals. Note that, owing to premultiplication, the power-law sections of the premultiplied PDFs are comparatively flattened, whereas an exaggeration occurs at the larger values of $(t_{p}\overline{u})/z$ (for a clear demonstration see figure \ref{fig:8} in Appendix \ref{app_A}). 

From figure \ref{fig:3}a, we notice that in highly convective stability the premultiplied PDFs of the positive and negative $T^{\prime}$ signals are clearly separated at larger values of $(t_{p}\overline{u})/z>4$ (shown by the divergence between the red and blue circles in figure \ref{fig:3}a). However, they agree with each other at the smaller values of $(t_{p}\overline{u})/z$. Using (\ref{t_p6}), we can conclude that this difference between positive and negative $T^{\prime}$ signals at the large values of $(t_{p}\overline{u})/z$ is related to the non-Gaussian characteristics of the temperature fluctuations, expressed by its skewness. \citet{chowdhuri2019revisiting} noted that the skewness and kurtosis of the temperature fluctuations are strongly non-Gaussian ($\approx$ 1.5 and 5 respectively) in the local free-convection limit ($-\zeta>$ 1). This observation is in agreement with the previous studies in the convective surface layer \citep{chu1996probability,liu2011probability,garai2013interaction,lyu2018high}. Our results show that this non-Gaussianity in the $T^{\prime}$ signal is only realised at large sizes of persistent temperature patterns, approximately greater than 4 times the measurement height. 

It is also interesting to note that, the total time fractions at these large sizes are governed by the negative fluctuation patterns alone (figure \ref{fig:3}a). \citet{katul1997turbulent_b} have shown that the difference in the PDFs between the positive and negative $T^{\prime}$ signals could be attributed to the asymmetry between the warm-updraft and cold-downdraft motions, under the assumption that the contributions from the counter-gradient quadrants could be ignored. Therefore, this finding is consistent with \citet{adrian1986turbulent}, where they showed from laboratory experiments that in highly-unstable conditions the temperature fluctuation patterns are governed by the more frequent cold-downdrafts bringing well-mixed air from aloft, interspersed with intermittent occurrences of the warm-updraft motions rising from the ground. 

However, this clear disparity between the premultiplied PDFs of positive and negative $T^{\prime}$ signals, gradually disappears as the near neutral stability is approached (figures \ref{fig:3}a to \ref{fig:3}f). This is congruous with the close to Gaussian characteristics of the $T^{\prime}$ signal in the near neutral stability \citep{chowdhuri2015relationship,chowdhuri2019evaluation}. Apart from that, for $u^{\prime}$ and $w^{\prime}$ signals no difference is observed between the premultiplied PDFs of positive and negative fluctuations, irrespective of the stability classes. The reason for this is the PDFs of $u^{\prime}$ and $w^{\prime}$ signals remain very nearly Gaussian for all the stability conditions in an atmospheric surface layer \citep{chu1996probability,chowdhuri2019revisiting}.   

In summary, for convective surface layer turbulence the persistence PDFs of the velocity and temperature fluctuations follow a power-law function up to a certain threshold size. This threshold size and the power-law exponents are not similar for the $u^{\prime}$, $w^{\prime}$, and $T^{\prime}$ signals. For the $u^{\prime}$ signals the threshold size is an order of magnitude larger than $z$, whereas for the $w^{\prime}$ signals it is approximately equal to $z$. On the other hand, the threshold size for the $T^{\prime}$ signals is somewhere in between of $u^{\prime}$ and $w^{\prime}$ signals. It is also remarkable to note that in a convective surface layer the non-Gaussian effects in the $T^{\prime}$ signals are only perceived at those sizes which are larger than this threshold size, where the power-law behaviour ceases to exist. Therefore, it is imperative to ask in convective surface layer turbulence,
\begin{enumerate}
    \item What statistical properties of velocity and temperature fluctuations give rise to the power-law behaviour punctuated with a cut-off in their persistence PDFs?
    \item How are those statistical properties connected with the structures of the turbulent flow?
\end{enumerate}
In the subsequent sections, we will explore these questions by analysing surrogate datasets and employing an alternate scaling of persistence time scales of velocity and temperature fluctuations.

\subsection{Scrutiny of persistence PDFs through surrogate data}
\label{surrogate_data}
The persistence PDFs are related to the inter-arrival time between the successive zero-crossings of a time series. If these zero-crossings were randomly located being independent of each other, we would have expected the persistence PDFs to follow a Poisson distribution, which is exponential in nature  \citep{majumdar1999persistence,santhanam2008return}. However, in a convective surface layer the persistence PDFs of the turbulent velocity and temperature fluctuations show a power-law structure. This indicates the zero-crossings of these signals are not independent but correlated with each other. Moreover, for the $T^{\prime}$ signals the power-law behaviour extends up to a certain streamwise size ($\approx 4z$) beyond which the non-Gaussian effects dominate the characteristics of the persistence PDFs. 

Therefore, to gain more insight regarding the statistical characteristics of the persistence PDFs of $u^{\prime}$, $w^{\prime}$, and $T^{\prime}$ signals in convective turbulence, we generated two different types of surrogate datasets, such as:
\begin{enumerate}
    \item The temporal correlation of the original time series is preserved by keeping the auto-correlation function unchanged, but modifying the PDF of the time series to be Gaussian.
    \item The temporal correlation of the original time series is destroyed by random shuffling, but keeping the PDF of the time series unchanged.
\end{enumerate}
The normal procedure to generate the first type of the surrogate dataset is taking the Fourier transform of the time series and keeping the Fourier amplitudes same but randomizing the associated Fourier phases, with inverse Fourier transform being applied to the modified Fourier coefficients \citep{poggi2009flume,lancaster2018surrogate}. Since the Fourier amplitudes of the surrogate dataset are kept intact during the process of Fourier phase-randomization, this procedure preserves the Fourier spectrum and hence the auto-correlation function of the time series. On the other hand, the second type of the surrogate dataset is generated by randomly permuting (shuffling) the original time series such that the temporal correlations are destroyed completely.

\subsubsection{Phase-randomization and randomization experiments}
For our purpose, we performed the phase-randomization and randomization experiments by varying the strength of the randomization, to investigate their gradual effects on the behaviour of the persistence PDFs. A similar procedure was suggested by \citet{maiwald2008surrogate} to investigate the effect of the non-linearity on a stochastic time series by increasing the strength of the phase-randomization in a step-wise manner. In this context, the randomization strength implies the percentage of the Fourier phases or the time series values which have been shuffled randomly to generate the surrogate datasets (see Appendix \ref{app_B} for further details). Figure \ref{fig:4} shows the typical results from these two experiments for the highly convective stability class. Note that, similar results have been found for the other five stability classes as well, which are not shown here. For the illustration purpose, the logarithmic persistence PDFs of positive and negative $T^{\prime}$ signals are shown at various randomization strengths from 0\% (original time series) to 100\% (completely randomized) for phase-randomization (figure \ref{fig:4}a) and randomization experiments (figure \ref{fig:4}b) respectively (see figures \ref{fig:s2} and \ref{fig:s3} in the supplementary material for the effect on $u^{\prime}$ and $w^{\prime}$ signals). The logarithmic representation is chosen, since in this representation the non-Gaussian characteristics of the $T^{\prime}$ signals are better revealed while preserving the power-law structure of the persistence PDFs (figures \ref{fig:3} and \ref{fig:8}). 

\begin{figure}[h]
\centering
\hspace*{-0.75in}
\includegraphics[width=1.25\textwidth]{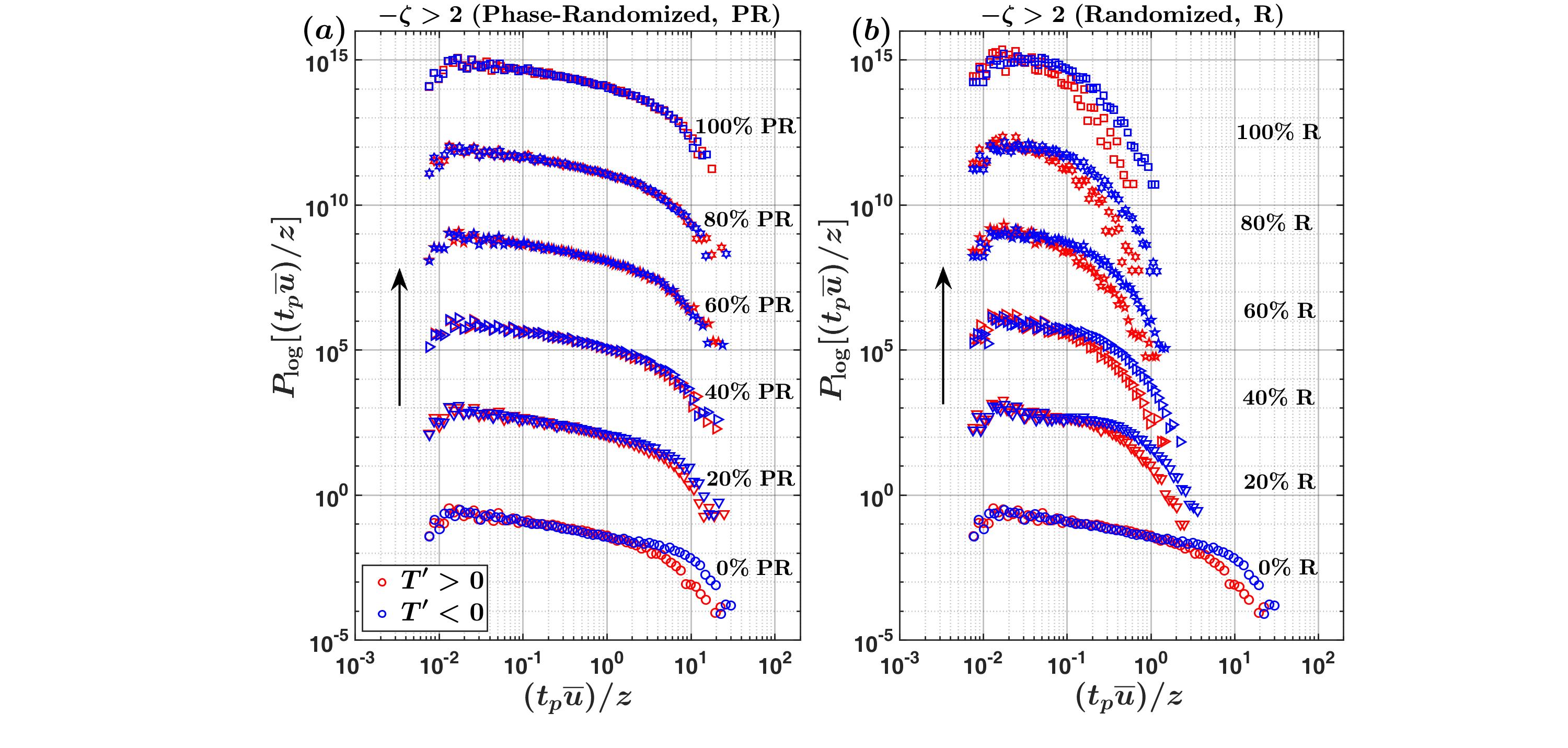}
  \caption{The logarithmic persistence PDFs of the normalized streamwise sizes $(t_{p}\overline{u})/z$ corresponding to the positive (red) and negative (blue) fluctuations in the temperature ($T^{\prime}$) signal from the highly convective stability class ($-\zeta>$ 2) are shown, for two sets of experiments such as: (a) Fourier phase-randomization (PR) and (b) temporal randomization (R). The original logarithmic persistence PDFs are shown at the bottom of both the panels and the rest are shifted vertically by three decades where the original temperature signals are gradually randomized either in their Fourier phases or temporal values, starting from 20 to 100\%.}
\label{fig:4}
\end{figure}

From figure \ref{fig:4}a we note that the power-law structure of the $T^{\prime}$ persistence PDFs are preserved for all the randomization strengths of the phase-randomization experiments, although the separation between the positive and negative $T^{\prime}$ persistence PDFs become indistinguishable at 20\% randomization strength. This is related to the fact that in phase-randomization experiments, the surrogate time series of temperature fluctuations approach an almost Gaussian distribution at even 20\% randomization strength (see figure \ref{fig:s4}a in the supplementary material). On the contrary, from figure \ref{fig:4}b we notice that the power-law structure of the $T^{\prime}$ persistence PDFs gradually disappear as the strength of the randomization is increased to 100\%. Nevertheless, in figure \ref{fig:4}b the difference between the positive and negative $T^{\prime}$ persistence PDFs remain preserved, irrespective of the randomization strength. The reason for this is, the non-Gaussian PDFs of the temperature fluctuations remain unchanged in the randomization experiments (see figure \ref{fig:s4}b in the supplementary material). Similar results are also obtained for the persistence PDFs of $u^{\prime}$ and $w^{\prime}$ signals (see figures \ref{fig:s2} and \ref{fig:s3} in the supplementary material).

\begin{figure}[h]
\centering
\hspace*{-0.75in}
\includegraphics[width=1.25\textwidth]{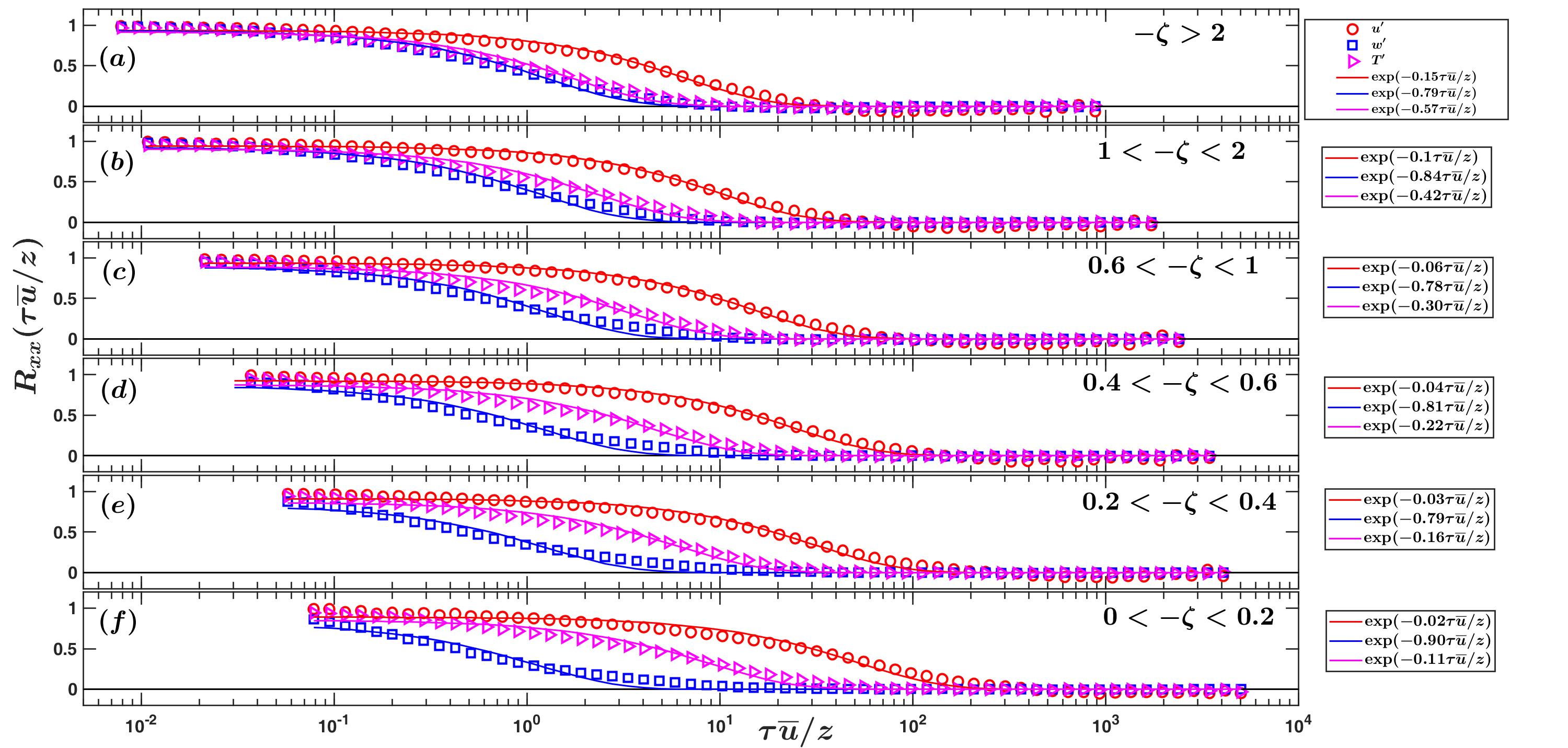}
  \caption{The auto-correlation functions are plotted against the normalized time lags ($\tau \overline{u}/z$) for $u^{\prime}$ (red circles), $w^{\prime}$ (blue squares), and $T^{\prime}$ (pink inverted triangles) signals for the six different stability classes as indicated at the right top most corner of each panel. The exponential functions are fitted to the auto-correlation functions to determine the integral length scale associated with $u^{\prime}$, $w^{\prime}$, and $T^{\prime}$ signals. The legends shown at the right of each panel describe the markers and the equations for the exponential fits.}
\label{fig:5}
\end{figure}

To summarize the results from figure \ref{fig:4}, we can conclude the power-law behaviour of the persistence PDFs is related to the temporal coherence in the time series. In a turbulent time series, the temporal coherence can be described by the integral time scale, defined as the time up to which the signal remains auto-correlated with itself \citep{tennekes1972first,kaimal1994atmospheric,wyngaard2010turbulence}.

\subsubsection{Auto-correlation functions and integral scales}
Figure \ref{fig:5} shows the auto-correlation functions ($R_{xx}(\tau)$, where $x$ can be either $u^{\prime}$, $w^{\prime}$, or $T^{\prime}$ signals and $\tau$ is the time lag) of the $u^{\prime}$, $w^{\prime}$, and $T^{\prime}$ signals, plotted against the lags for the six different stability classes. Note that in figure \ref{fig:5}, the time lags are converted to a streamwise length ($\tau \overline{u}$) by using Taylor's frozen turbulence hypothesis and normalized with $z$. Apart from that, the auto-correlation functions are ensemble averaged over the set of 30-min runs from a particular stability class (see table \ref{tab:1}) to ensure the results are statistically robust. These auto-correlations can be represented with an exponentially decaying function of the form,
\begin{equation}
R_{xx}(\frac{\tau \overline{u}}{z})=\exp(-K \frac{\tau \overline{u}}{z}) \implies \exp(-\frac{\ell_{x}}{\Lambda_{x}}),
\label{auto_corr}
\end{equation}
where $x$ can be either $u^{\prime}$, $w^{\prime}$, or $T^{\prime}$, $\ell_{x}=\tau \overline{u}$, and $\Lambda_{x}=z/K$, known as the integral length scale \citep{kaimal1994atmospheric}.

From figure \ref{fig:5}a, we note that the coefficients $K$ for the $T^{\prime}$ and $w^{\prime}$ signals are closer to each other ($K=0.57$ and $0.79$ respectively), whereas the $K$ value for the $u^{\prime}$ signal remains much lower ($K=0.15$). Using (\ref{auto_corr}), this implies in the highly convective stability the integral length scales of the $T^{\prime}$ and $w^{\prime}$ signals ($\Lambda_{x}=z/K$) almost coincide with each other, while the integral length scale of the $u^{\prime}$ signal remains the largest. However, as the stability changes from highly convective to near neutral the coefficients $K$ of the $T^{\prime}$ signals become closer to the $u^{\prime}$ signals, implying the $\Lambda_{T}$ values approach $\Lambda_{u}$ (figure \ref{fig:5}a to \ref{fig:5}f). On the other hand, the $\Lambda_{w}$ values remain approximately equal to $z$ irrespective of the stability classes ($K=0.79$ to $0.90$). 

These results are in accord with the previous studies in surface layer turbulence, where it has been noted that in highly convective stability the $\Lambda_{T}$ values approach $\Lambda_{w}$, whereas in the near neutral stability they are much closer to $\Lambda_{u}$ \citep{kader1989spatial,katul1997energy,li2012mean}. Apart from that, the approximate equality of $\Lambda_{w}$ with $z$ supports the argument that the eddies which contribute to the $w^{\prime}$ signal are attached to the ground \citep{mcnaughton2007scaling,banerjee2013logarithmic,banerjee2015revisiting,chowdhuri2019empirical}. If these integral length scales obtained from figure \ref{fig:5} ($\Lambda_{u}$, $\Lambda_{w}$, and $\Lambda_{T}$) are replaced in the persistence length scales ($t_{p}\overline{u}$) in figure \ref{fig:3}, we notice that the deviation from the power-law occurs at those scales larger than the integral scales, for the $u^{\prime}$, $w^{\prime}$, and $T^{\prime}$ signals. The integral length scales are comparable to the peak wavelength of the turbulence energy spectrum \citep{kaimal1994atmospheric}. This is demonstrated in figure \ref{fig:s5} (see supplementary material) for the highly convective stability class ($-\zeta>2$), where the scaled peak wavenumbers ($\kappa z$, where $\kappa$ is the streamwise wavenumber) of the $u^{\prime}$, $w^{\prime}$, and $T^{\prime}$ spectra match with $z/\Lambda_{x}$ ($x$ can be either $u^{\prime}$, $w^{\prime}$, or $T^{\prime}$ signals). Therefore, it indicates the power-law behaviour of the persistence PDFs are connected to the eddies from the inertial subrange of the turbulence spectrum (sizes smaller than the integral scales). On the other hand, the departure from the power-law behaviour in the persistence PDFs occurs at those scales comparable to the energy containing scale of eddies (sizes larger than the integral scales). To further explore this connection with the turbulent flow structures, it is appropriate to represent the persistence PDFs by scaling their persistence times with the integral scales. We present the results regarding those in the subsequent section.   

\subsection{Scaling the persistence time by the integral scales}
\label{scaling}
We begin with discussing the logarithmic persistence PDFs, since in that representation the disparity between the PDFs corresponding to the positive and negative fluctuations at larger persistence scales are highlighted more clearly (see figure \ref{fig:3} in Section \ref{persistence_PDFs}). The persistence time $t_{p}$ of $u^{\prime}$, $w^{\prime}$, or $T^{\prime}$ signals are scaled with the integral time scale ($\Gamma$) before computing the logarithmic persistence PDFs. Note that, from the application of Taylor's frozen turbulence hypothesis this is equivalent to scaling the persistence length with integral scales $\Lambda$. 

\subsubsection{Logarithmic persistence PDFs}
\begin{figure}[h]
\centering
\hspace*{-0.75in}
\includegraphics[width=1.25\textwidth]{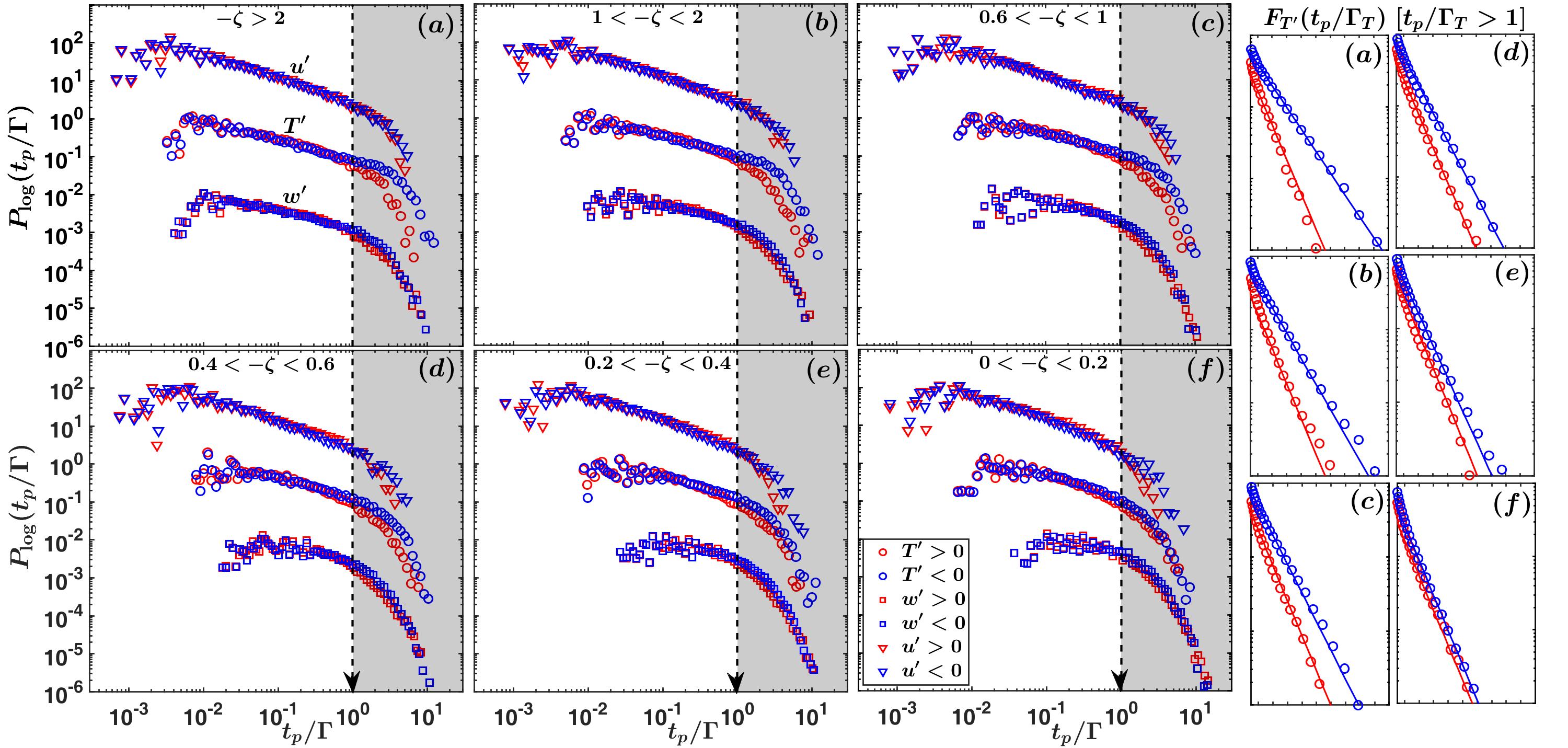}
  \caption{The logarithmic persistence PDFs are shown for the persistence times normalized by the integral time scales ($\Gamma$) associated with $w^{\prime}$ (squares), $T^{\prime}$ (circles), and $u^{\prime}$ (inverted triangles) signals. The logarithmic PDFs are shifted vertically for the visualization purpose. The panels on the right show the cumulative distribution functions of the $T^{\prime}$ signals ($F_{T^{\prime}}(t_{p}/\Gamma_{T})$) plotted for $t_{p}/\Gamma_{T}>1$ (marked as the grey regions on the left panels) on a log-linear plot. This representation is chosen to determine the slope of the exponential functions (represented as straight lines, see (\ref{cdf_1})) fitted separately for the positive and negative temperature fluctuations persisting for times larger than the integral time scales. The description of the markers are shown in the legend, placed in a corner of the bottom panel (f) at the left hand side.}
\label{fig:6}
\end{figure}

Figure \ref{fig:6} shows the logarithmic persistence PDFs with the persistence times normalized by $\Gamma$ corresponding to $u^{\prime}$, $w^{\prime}$, or $T^{\prime}$ signals, for all the six stability classes (table \ref{tab:1}). For all the three signals in figure \ref{fig:6}, a bend could be clearly observed in the logarithmic persistence PDFs at the time scales larger than the integral scales, indicating the clear deviation from the power-law. Apart from that, one can notice from figure \ref{fig:6}a (highly convective stability) that the difference between the positive and negative persistence PDFs for the $T^{\prime}$ signals occurs exactly at those time scales which are larger than the integral time scales of temperature (see the grey marked region in figure \ref{fig:6}a). This discrepancy gradually disappears in near neutral stability (see figures \ref{fig:6}a to \ref{fig:6}f), implying the non-Gaussian characteristics of the $T^{\prime}$ signal are definitely related to the energy containing scales of motions. 

However, the persistence PDFs deviate from the power-law form at scales larger than the integral scales. \citet{chamecki2013persistence} showed for canopy surface layer turbulence, the persistence PDFs of velocity fluctuations at large time scales behave exponentially, the hallmark of a random Poisson type process. \citet{cava2012role} hypothesized this as a consequence of random deformation of the coherent structures, giving rise to several sub-structures with independent arrival times. Be that as it may, \citet{chamecki2013persistence} also commented that this exponential nature of the persistence PDFs could be better studied by considering the cumulative distribution functions (CDF), given their smoothed nature at large time scales. 

The CDF [$F(t_{p}/\Gamma)$] is defined as,
\begin{equation}
F(t_{p}/\Gamma)=\int_{{(t_{p}/\Gamma)}_{\rm max}}^{(t_{p}/\Gamma)} P(t_{p}/\Gamma) \, d{(t_{p}/\Gamma)},
\label{cdf}
\end{equation}
which denotes the probability that the normalized persistence time has a value smaller than or equal to $t_{p}/\Gamma$. From (\ref{cdf}), it follows that for an exponential distribution (Poisson process) both the PDFs and CDFs have the same form, albeit with different proportionality constants. Additionally, the CDFs also have an inherent advantage of being bin-independent with a smooth convergence towards 1 \citep{newman2005power,white2008estimating}. At the right hand side of figure \ref{fig:6} (see the side panels), the CDFs corresponding to the positive and negative $T^{\prime}$ signals are shown for the time scales larger than the integral scale $\Gamma_{T}$ [$(t_{p}/\Gamma_{T})>1$] in a log-linear co-ordinate system, corresponding to all the six stability classes (the CDFs of $u^{\prime}$, $w^{\prime}$, and $T^{\prime}$ signals for the whole range of $t_{p}/\Gamma$ are provided in figure \ref{fig:s6}). The exponential decay of the CDFs,
\begin{equation}
F(t_{p}/\Gamma) \propto \exp \Big[-\lambda (t_{p}/\Gamma)\Big],
\label{cdf_1}
\end{equation}
in such plots would appear as a straight line with the slope of $\lambda$. 

For the highly convective stability (panel (a) at the right hand side of figure \ref{fig:6}), we note that the $\lambda$ values obtained from the slopes of the straight line fits for negative and positive $T^{\prime}$ are 0.7 and 1.1 respectively. It implies the mean time scales ($\Gamma/\lambda$) related to the long persistent events of negative $T^{\prime}$ are almost twice the integral scale of temperature, whereas for the positive $T^{\prime}$ they are almost equal to the integral scale. Furthermore, with the change in stability from highly convective to near neutral, the $\lambda$ values for the long events of positive $T^{\prime}$ remain fixed at 1.1, but for the negative $T^{\prime}$ these values gradually increase from 0.7 to 1.1 at near neutral stability (see the side panels (a) to (f) in figure \ref{fig:6}). This indicates the mean time scales of the long persistent events of negative temperature fluctuations become closer to the integral scale of temperature as the near neutral stability is approached, while the same remains unchanged for the positive fluctuation patterns.

Since these long events are associated with the energy containing scales (see Section \ref{surrogate_data}), the disparity in the mean time scales (or length scales from Taylor's frozen turbulence hypothesis) between the warm and cold events could be explained from the coherent structure perspective. The large-eddy simulation (LES) results of \citet{khanna1998three} and \citet{salesky2017nature} indicate that in highly convective stability the vertical velocity and temperature fluctuations display a cellular organization pattern. This explains the closeness of the integral scales of $w^{\prime}$ and $T^{\prime}$ in highly convective stability (figure \ref{fig:5}a). Apart from that, their LES simulations also show that the warm-updraft motions are concentrated at the cell edges, while the cold-downdraft motions occupy a larger area at the cell centre. This geometrical asymmetry between the warm-updraft and cold-downdraft motions might be the reason why the long warm events have mean length scales comparable to the integral scales, while the scales of the long cold events are almost twice of that. However, in near neutral conditions the warm and cold fluids are positioned on the long streaky patterns of streamwise fluctuations having similar streamwise lengths, comparable to the integral scales \citep{khanna1998three,salesky2017nature}. Given the closeness of $\Lambda_{T}$ and $\Lambda_{u}$ in the near neutral stability (figure \ref{fig:5}f), this elucidates why in such conditions an equivalence is observed between the mean length scales of long warm and cold events. 

\subsubsection{Persistence PDFs}

\begin{figure}[h]
\centering
\hspace*{-0.7in}
\includegraphics[width=1.25\textwidth]{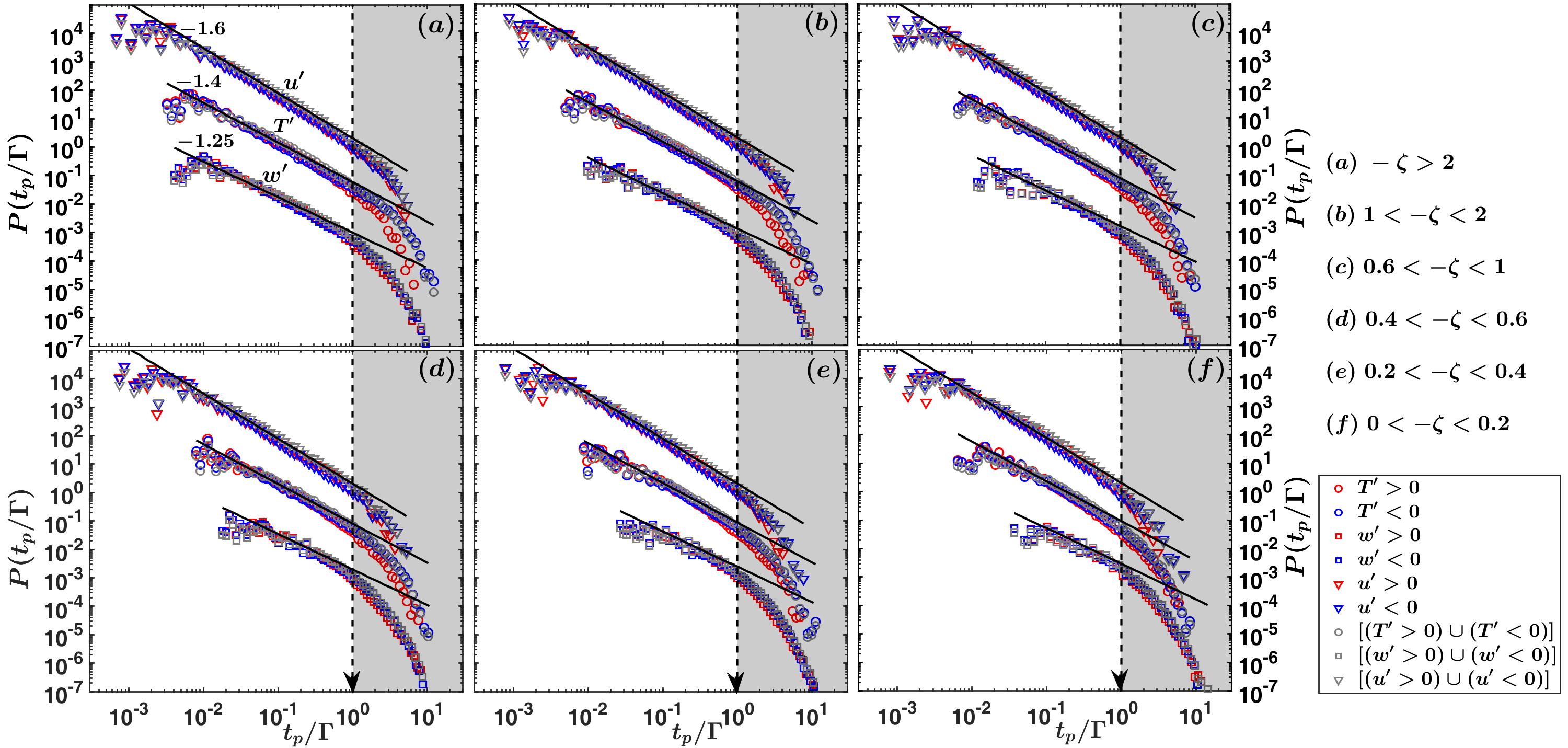}
  \caption{The persistence PDFs are shown for the persistence times normalized by the integral time scales ($\Gamma$) associated with $w^{\prime}$ (squares), $T^{\prime}$ (circles), and $u^{\prime}$ (inverted triangles) signals. The thick black lines on all the panels show the best fit power-laws with their respective slopes being mentioned on the panel (a). The regions on all the panels corresponding to $t_{p}/\Gamma>1$ are marked in grey. The description of the markers are shown in the legend, placed at the right bottom corner.}
\label{fig:7}
\end{figure}

So far in figure \ref{fig:6}, we have focused on the characteristics of the long persistent events larger than the integral scales (associated with energy containing eddies) by investigating the logarithmic representation of the persistence PDFs. However, it would also be advantageous to focus on the power-law structure of the persistence PDFs at scales smaller than the integral scales (associated with the eddies from the inertial subrange). Since the power-law behaviour is best represented in the original PDFs, figure \ref{fig:7} shows the persistence PDFs of $u^{\prime}$, $w^{\prime}$, and $T^{\prime}$ signals with $t_{p}$ scaled with $\Gamma$. 

From figure \ref{fig:7}, one can notice the cut-off scale, where the deviation from the power-law behaviour begins, is located almost exactly at the integral scale $\Gamma$ for all the three signals. Apart from that, compared to figure \ref{fig:2}, the cut-off behaviour of the power-law is quite sharp in figure \ref{fig:7} due to the correct scaling of $t_{p}$ with $\Gamma$. Therefore, in figure \ref{fig:7}a (highly convective stability) the exponents of the power-law functions are determined by performing a linear regression for the range 0.01 $\leq t_{p}/\Gamma \leq$ 1 on the log-log plots. The slopes (exponents) of the best fit lines are found to be $-1.6$, $-1.4$, and $-1.25$ for $u^{\prime}$, $w^{\prime}$, and $T^{\prime}$ signals respectively, with $R^{2}$ values more than 0.98. Note that, the exponent 1.4 for the $T^{\prime}$ signal is very close to 1.37 as reported by \citet{bershadskii2004clusterization} from their turbulent convection experiments. To assess the effect of stability on the power-law behaviour, the exponents obtained from the highly convective stability are compared with the other stability classes. The following features emerge, such as:
\begin{enumerate}
    \item For the $u^{\prime}$ signal, no change in the power-law behaviour is observed with stability.
    \item For the $T^{\prime}$ signal, no discernible change can be observed, although at near neutral stability a slight departure from $-1.4$ exponent might be possible.
    \item For the $w^{\prime}$ signal, the extent of the power-law behaviour gradually decreases with stability.
\end{enumerate}

Since the power-law structure in the persistence PDFs is associated with the $t_{p}$ values smaller than the integral time scales, it is thus representative of the eddies from the inertial subrange of the turbulence spectrum. This observation lends considerable support to the hypothesis of \citet{yee1995measurements} and \citet{cava2012role} where they connected this power law behaviour in the persistence PDFs with self-similar Richardson cascading mechanism, given the implied scale invariance associated with power-law distributions \citep{newman2005power,verma2006universal}. Accordingly, it also explains the reduction in the extension of the power-law behaviour for the $w^{\prime}$ signal, associated with the near neutral stability. From table \ref{tab:1}, it is clear that the near neutral stability class ($0<-\zeta<0.2$) corresponds only to the lowest three levels of the SLTEST experiment ($z=$ 1.4, 2.1, 3 m). The inertial subrange of the $w$ spectra starts approximately at $\kappa z=1$, considering $\Lambda_{w}\approx z$. If this is converted to frequency ($f=\overline{u}/2 \pi z$), then we would obtain the threshold at $\approx$ 1 Hz (assuming  $\overline{u} \approx$ 10 m s$^{-1}$ and $z=2$ m), indicating only a decade wide subrange being resolved at 20-Hz sampling rate. Therefore, the decrease in the power-law range for the $w^{\prime}$ signal could be attributed to insufficient sampling of small scale eddies at 20-Hz frequency for the lowest three SLTEST levels.

\subsubsection{Physical explanation of persistence exponents}
In general, the exponents of the power-laws in persistence PDFs are non-trivial and difficult to compute analytically except for simple systems such as fractional Brownian motions \citep{majumdar1999persistence,aurzada2013persistence,aurzada2015persistence}. Recently, to explain these power-law exponents, \citet{cava2009effects} and \citet{cava2012role} have proposed an ambitious connection with the self-organized criticality (SOC) observed in the sandpile model of \citet{bak59self,bak1988self}. This is inspired by the results from \citet{sreenivasan2004multiscale} where such a connection was proposed for turbulent convection. By using the results from \citet{jensen19891} and \citet{bershadskii2004clusterization}, \citet{cava2012role} connects this power-law exponents as,
\begin{equation}
m=3-\gamma-\frac{\mu}{2},
\label{soc}
\end{equation}
where $m$ is the spectral slope of the telegraphic approximated time series, $\gamma$ is the power-law exponent, and $\mu$ is the intermittency coefficient. \citet{sreenivasan2006clustering} found $m$ to be equal to 4/3, as opposed to the 5/3 spectral slope in the inertial subrange. If we accept this premise of SOC and assume $m$ to be the same for $u^{\prime}$, $w^{\prime}$, and $T^{\prime}$ signals, then from (\ref{soc}) we can comment that the intermittency coefficients ($\mu$) are larger for the $w^{\prime}$ and $T^{\prime}$ signals ($\gamma=$ 1.25 and 1.4 respectively) compared to the $u^{\prime}$ signal ($\gamma=$ 1.6). This implies in a convective surface layer, the $w^{\prime}$ and $T^{\prime}$ signals are more intermittent in nature than the $u^{\prime}$ signal. 

However, an alternate theoretical framework also exists where the persistence PDFs are connected to the fractal dimensions of small-scale turbulence \citep{catrakis1996scale,dimotakis1999turbulence,catrakis2000distribution}. In retrospect, this alternate framework had been proposed to provide a new methodology to compute the fractal dimensions, without relying on the box-counting methods being used in the studies by \citet{sreenivasan1986fractal} and \citet{scotti1995fractal} (see \citet{sreenivasan1991fractals} and \citet{catrakis2008multiscale} for a review). Nonetheless, at present, both of these frameworks based on SOC and fractals seem plausible to connect these power-law exponents with the physical nature of small-scale turbulence, but further research is required to assess their viability. We present our conclusions in the next section. 

\section{Conclusion}
\label{conclusion}
We report the statistical scaling properties of the persistence PDFs of turbulent fluctuations in streamwise and vertical velocity ($u^{\prime}$ and $w^{\prime}$) and temperature ($T^{\prime}$) from the SLTEST experimental dataset, in a convective surface layer. The important results from this study can be summarized as:
\begin{enumerate}
    \item The persistence PDFs of $u^{\prime}$, $w^{\prime}$, and $T^{\prime}$ signals display a power-law behaviour up to a certain threshold scale, punctuated with an exponential cut-off. The power-law exponents are 1.6, 1.25, and 1.4 for the $u^{\prime}$, $w^{\prime}$, and $T^{\prime}$ signals respectively, with no significant change being observed with stability.
    \item From randomization experiments, it is shown that this power-law behaviour of the persistence PDFs is linked to the temporal coherence in the time series. This temporal coherence is represented by the integral scales, computed from the auto-correlation functions. 
    \item By normalizing the persistence time scales with the integral scales, it is found that the power-law behaviour in the PDFs is related to those scales which are smaller than the integral scales. Since power-laws are synonymous with scale-invariance, it implies the effect of self-similar eddy cascading mechanism (Richardson cascade) on the persistence PDFs.
    \item A premultiplied form of the persistence PDF is used to demonstrate that the non-Gaussian effects of the temperature fluctuations act only at those scales which are larger than the integral scales. This indicates the non-Gaussian characteristics of the temperature fluctuations are associated with the energy containing scales of motions.
    \item From the exponential fits in the persistence CDFs, it is illustrated that the mean time scales of the negative $T^{\prime}$ events persisting longer than the integral scales are approximately twice the size of the integral scales in highly convective conditions. However, this mean time scale gradually decreases to almost being equal to the integral scale in the near neutral stability.
    \item On the other hand, for the long positive $T^{\prime}$ events, the mean time scales remain roughly equal to the integral scales, irrespective of stability. This discrepancy with the negative $T^{\prime}$ events is interpreted to be associated with the change in the topology of the coherent structures from cellular convection patterns in highly convective conditions to horizontal streaks in near neutral stability.  
    \end{enumerate}
Note that, this study is the first of its kind for a convective surface layer turbulence, where the entire focus has been on establishing the statistical characteristics of the persistence PDFs. We have convincingly demonstrated that the persistence PDFs of velocity and temperature fluctuations in a convective surface layer follow a power-law distribution followed by an exponential cut-off. Subsequently, by scaling the persistence time scales with the integral scales, this statistical property of the persistence PDFs has been associated with the turbulent structures in a convective flow. Apparently, this scaling has been formulated by generating surrogate datasets which preserve or destroy the temporal correlations in a turbulent signal. 

Apart from that, it is also important to emphasize that the findings obtained from this study have direct bearings on the quadrant behaviour of the turbulent momentum and heat fluxes. This is because the switching patterns from one quadrant to the other are related to the combination of the zero-crossings in the component signals which construct the fluxes ($u^{\prime}$ and $w^{\prime}$ or $w^{\prime}$ and $T^{\prime}$). Since persistence PDFs are related to the time spent between the zero-crossings in a signal, the future research questions are:
\begin{enumerate}
    \item How the persistence PDFs of the heat and momentum fluxes are related to the persistence PDFs of the velocity and temperature fluctuations?
    \item How much of the flux variation can be described by the persistence properties of the component signals?
    \end{enumerate}

\section*{Acknowledgements}
Indian Institute of Tropical Meteorology (IITM) is an autonomous institute fully funded by the Ministry of Earth Sciences, Government of India. T. Banerjee acknowledges the funding support from the University of California Laboratory Fees Research Program funded by the UC Office of the President (UCOP), grant ID LFR-20-653572. Additional support was provided by the new faculty start up grant provided by the Department of Civil and Environmental Engineering, and the Henry Samueli School of Engineering, University of California, Irvine. Dr. Kalm\'{a}r-Nagy acknowledges funding from the Higher Education Excellence Program of the Ministry of Human Capacities in the frame of the Water Science \& Disaster Prevention research area of Budapest University of Technology and Economics (BME FIKP-V\'{I}Z) and by the National Research, Development, and Innovation Fund (TUDFO/51757/2019-ITM, Thematic Excellence Program). The authors would also like to thank Dr. KG McNaughton for providing them the SLTEST dataset. The computer codes used in this study are available to all the
researchers by contacting the authors.  

\appendix
\appendixpage
\begin{appendices}
\section{The effect of binning on persistence PDFs}
\label{app_A}

\begin{figure}[h]
\centering
\hspace*{-3in}
\includegraphics[width=1.7\textwidth]{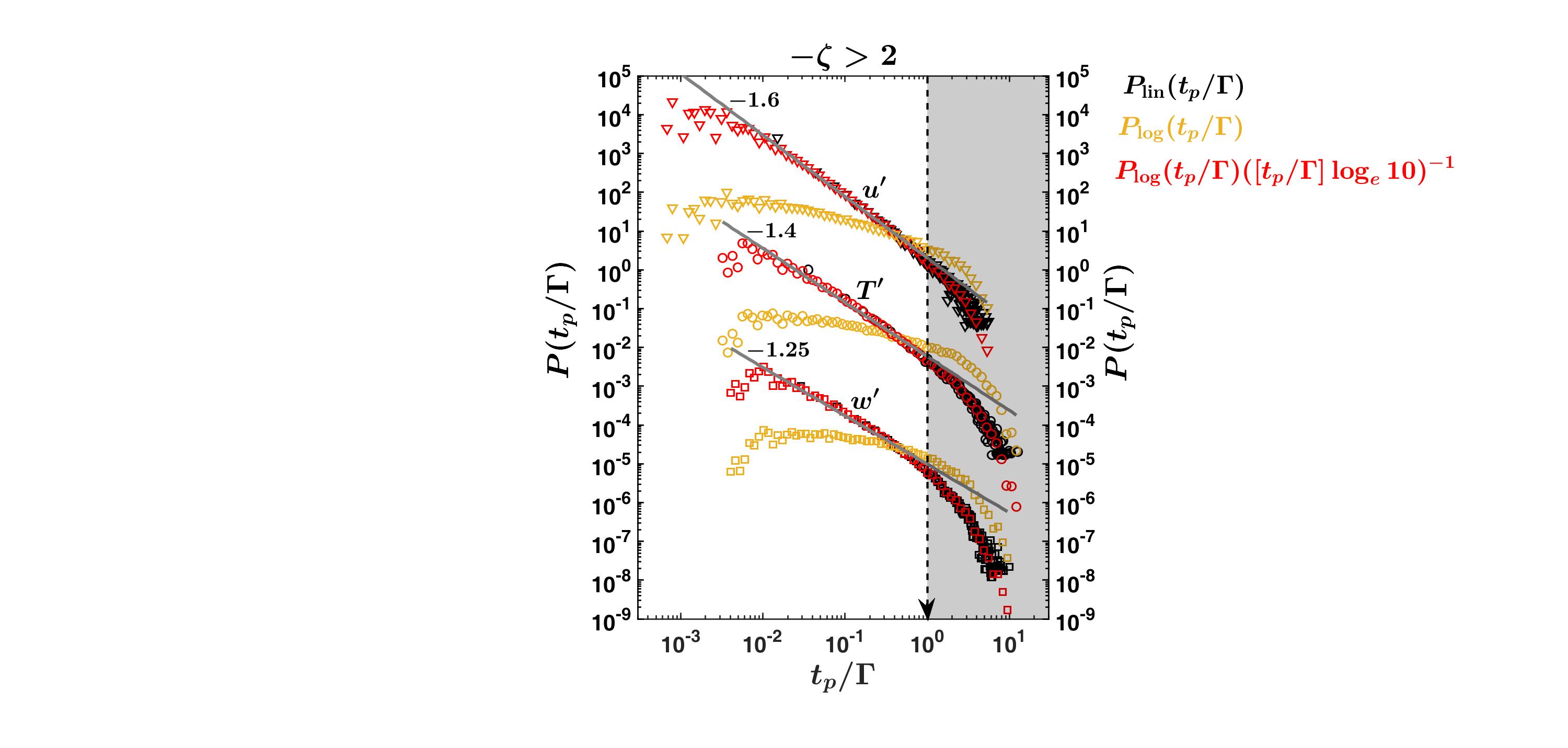}
\vspace{-10mm}
  \caption{An example is provided from the highly convective stability class ($-\zeta>2$) to illustrate the effect of linear and logarithmic binning on the persistence PDFs of $w^{\prime}$ (squares), $T^{\prime}$ (circles), and $u^{\prime}$ (inverted triangles) signals, considering both the positive and negative fluctuations. The persistence times are normalized by the integral time scales ($\Gamma$) associated with $w^{\prime}$, $T^{\prime}$, and $u^{\prime}$ signals. The black coloured markers indicate the persistence PDFs constructed from linear binning, whereas the orange coloured markers denote the same but from logarithmic binning. To convert from logarithmic to linear space, the logarithmic persistence PDFs are premultiplied with the factor ${([t_{p}/\Gamma]\log_{e}10)}^{-1}$ obtained from (\ref{pdf_conversion}) and are shown as the red coloured markers. Note that all the three PDFs for $w^{\prime}$, $T^{\prime}$, and $u^{\prime}$ signals are shifted vertically by three decades for visualization purpose. The thick grey lines show the same power-laws as in figures \ref{fig:2} and \ref{fig:7}.}
\label{fig:8}
\end{figure}

The common method for computing the PDFs of a stochastic variable is to bin the data in the linear space and computing the fraction of samples lying within each bin, divided by the bin-width \citep{panofsky1958some,lumley2007stochastic}. However, in the context of a stochastic variable which varies over a substantial range, sometimes many orders of magnitude, the linear binning may not be a good strategy. The PDFs obtained from linear binning for such variables are usually dominated by a few high probability bins at the lower range and an overwhelming amount of many low probability bins at the higher range, as illustrated by \citet{dorval2011estimating}. A better strategy is to use logarithmic binning for stochastic variables which have positively skewed distributions, such as having high densities at the lower range and low densities at the large range of values \citep{dorval2011estimating}. Intuitively, we may expect the same with the persistence time, having large occurrences associated with smaller time scales, with occasional incidences persisting for a long time. 

In the logarithmic binning exercise, a logarithmic transformation is applied to the associated variable and then the fractions of samples are computed in each logarithmic bin, divided by the bin-width in the logarithmic space to compute the PDFs. The logarithmic binning has been used extensively to deduce the characteristics of the distributions associated with neuronal inter-spike intervals in biophysical signals \citep{selinger2007methods,dorval2011estimating}, to estimate the size distribution of the avalanches associated with sandpile models \citep{christensen2005complexity}, identifying the l\'evy flight patterns in animal displacements \citep{benhamou2007many,sims2007minimizing}, and in many other practical cases (see \citet{newman2005power} and the references therein for a detailed review on this topic). 

However, the logarithmic bin-width is variable in the linear space and increases as the values increase, thus making the number of samples in each logarithmic bin dependent on the linear width of the bin \citep{white2008estimating}. By employing the change of variables technique, \citet{dorval2011estimating} showed that the PDFs constructed in the linear ($P_{\rm lin}$) and logarithmic ($P_{\rm log}$) spaces are related to each other as,
\begin{equation}
P_{\rm log}(x)=(\log_{e}10)[xP_{\rm lin}(x)],
\label{pdf_relation}
\end{equation}
where $x$ is the associated stochastic variable and the factor $\log_{e}10$ emerges as the bins of $x$ are constructed in the powers of 10. Note that, in this study we also used the bins in the power of 10 while constructing the persistence PDFs. Since the PDFs should be related to constant bin-width in the linear space \citep{panofsky1958some,sims2007minimizing}, we can rearrange (\ref{pdf_relation}) to obtain the actual PDFs from the logarithmic PDFs as,
\begin{equation}
P(x)=[P_{\rm log}(x)]{(x\log_{e}10)}^{-1}.
\label{pdf_conversion}
\end{equation}
An alternate way to obtain the actual PDFs is to divide the fraction of samples within each logarithmic bin by the equivalent linear width in the logarithmic space, also known as normalized logarithmic binning \citep{white2008estimating}. However, both of these methods yield the same result, since the PDFs obtained from the normalized logarithmic binning are equivalent to (\ref{pdf_conversion}). 

In figure \ref{fig:8}, we show an example of persistence PDFs for $u^{\prime}$, $w^{\prime}$, and $T^{\prime}$ signals (combining both the positive and negative fluctuations) from a highly convective stability class ($-\zeta>2$) to illustrate the effect of linear and logarithmic binning strategies. Before constructing the PDFs, the persistence time $t_{p}$ is normalized by the integral scale $\Gamma$ associated with $u^{\prime}$, $w^{\prime}$, and $T^{\prime}$ signals, due to the reasons as described in Section \ref{results}. We note that, the linear PDFs and the PDFs described by (\ref{pdf_conversion}) are equivalent to each other, although the PDFs obtained from (\ref{pdf_conversion}) display a broader range of power-law behaviour and substantially less noisy than the linear PDFs in the larger range of $t_{p}/\Gamma$ values. On the other hand, the logarithmic PDFs are premultiplied PDFs as shown in (\ref{pdf_relation}). Therefore, compared to the actual PDFs, the logarithmic PDFs exhibit a flatter slope in the smaller range of $t_{p}/\Gamma$ values, concomitant with a slower decrease in the larger range. 

\section{Description of phase-randomization and randomization experiments}
\label{app_B}
To perform the phase-randomization experiment at varying strength of randomization, we first choose a 30-min time series of $u^{\prime}$, $w^{\prime}$, or $T^{\prime}$ signals (having 36000 points at sampling frequency of 20-Hz) from a particular stability class, as outlined in table \ref{tab:1}. The step-by-step methodology is provided below, such as:
\begin{enumerate}
    \item The Fourier transformation of the 30-min time series is performed, generating 36000 complex Fourier coefficients. Out of the 36000 coefficients, half of these are the complex conjugates of each other, implying the need to only consider the Fourier phases of the first 18000 coefficients.
    \item These 18000 Fourier phases are separated into two parts along their midpoint at the 9000-th point. From the left half of the 9000 Fourier phases an $(\frac{x}{2})$\% of the phases are randomly chosen, with the same being repeated for the right half as well.
    \item These $(\frac{x}{2})$\% of the randomly chosen phases from the left and right halves are shuffled with each other. Therefore, total $x$\% Fourier phases of the 18000 coefficients are scrambled in this procedure while keeping their amplitudes intact. The revised Fourier coefficients are computed by using the same amplitude but the scrambled phases, and then their complex conjugates are taken for the rest of the 18000 points. 
    \item The inverse Fourier transform is undertaken for these revised Fourier coefficients, to generate the phase-randomized time series at an $x$\% randomization strength. The randomization strength implies the percentage of the Fourier phases which have been shuffled randomly to generate the surrogate dataset.
\end{enumerate}
Note that, the aforementioned procedure can be adopted for the randomization experiment as well. To generate a randomized dataset at an $x$\% randomization strength, $(\frac{x}{2})$\% of the time series values are randomly shuffled between the left and right halves, along the midpoint of the time series at 18000-th point. This type of shuffling procedure is known as Hurst’s card simulation, by drawing analogy with the shuffling patterns of a standard deck employed in card playing games \citep{strecker2004fractional}.  
\end{appendices}

\bibliographystyle{apalike}  
\bibliography{references}
\clearpage

\section*{Supplementary material}
\renewcommand{\thefigure}{S\arabic{figure}}
\setcounter{figure}{0}

\begin{figure}[h]
\centering
\vspace*{0.5in}
\hspace*{-0.75in}
\includegraphics[width=1.25\textwidth]{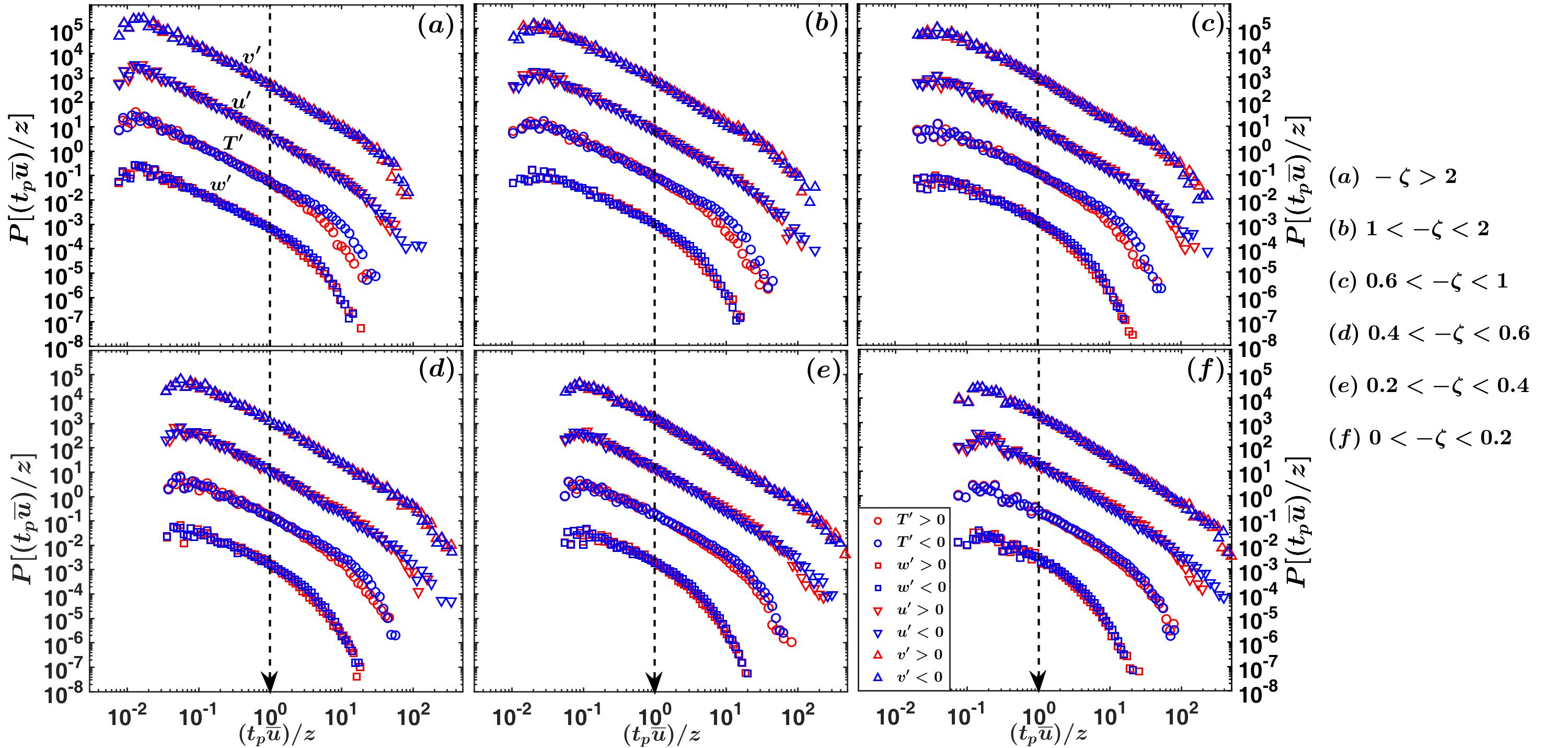}
 \caption{Same as in figure \ref{fig:2}, but for the cross-stream positive and negative fluctuations ($v^{\prime}$) included.}
\label{fig:s1}
\end{figure}

\begin{figure}[h]
\centering
\vspace*{0.5in}
\hspace*{-0.75in}
\includegraphics[width=1.25\textwidth]{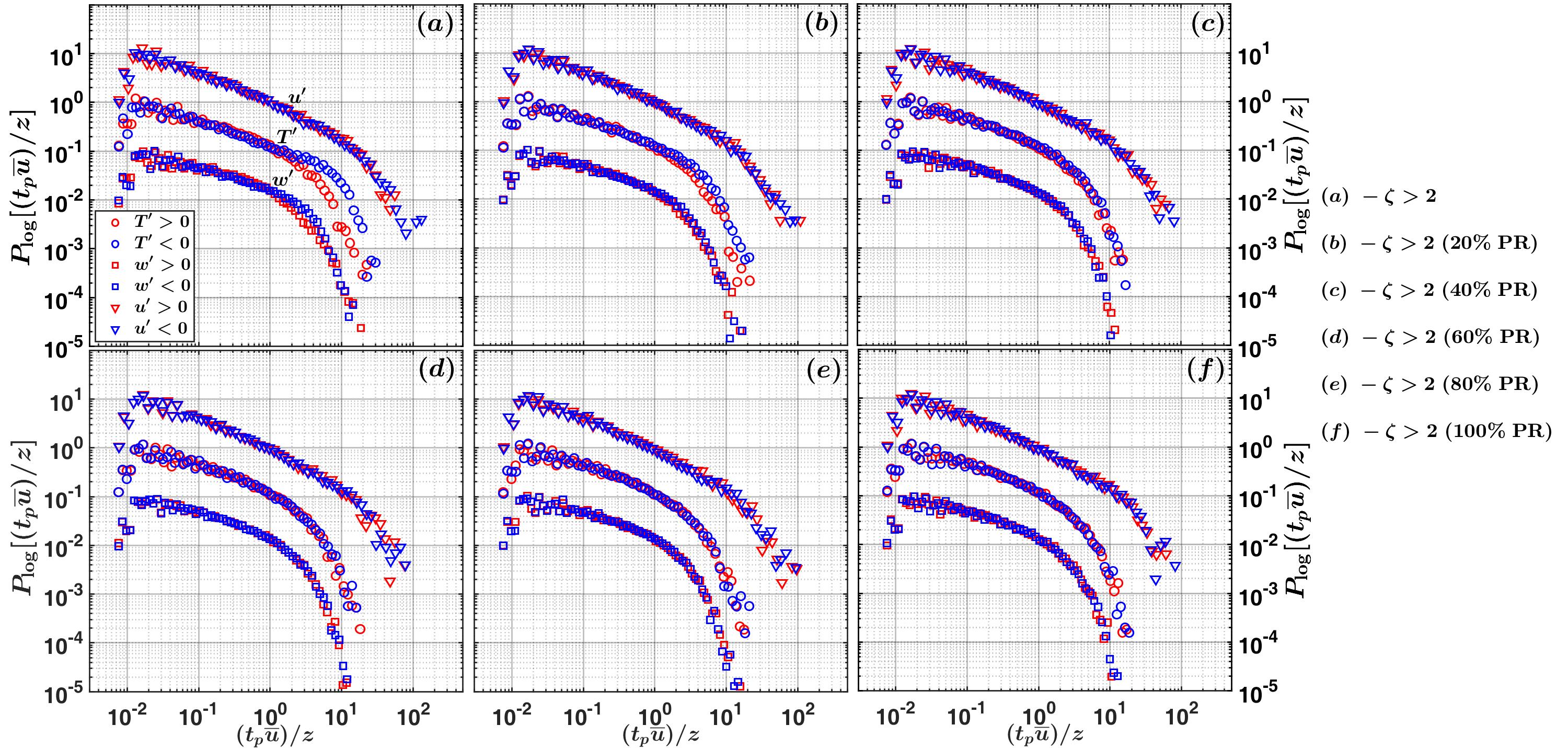}
 \caption{Same as in figure \ref{fig:4}, but for the positive and negative fluctuations in $u^{\prime}$ and $w^{\prime}$ signals also included from the highly convective stability class ($-\zeta$>2), during the phase-randomization experiment. The panels show the signals are gradually randomized in their Fourier phases, beginning from the original signals to 100\% phase-randomized signals, in the increments of 20\%.}
\label{fig:s2}
\end{figure}

\begin{figure}[h]
\centering
\vspace*{0.5in}
\hspace*{-0.75in}
\includegraphics[width=1.25\textwidth]{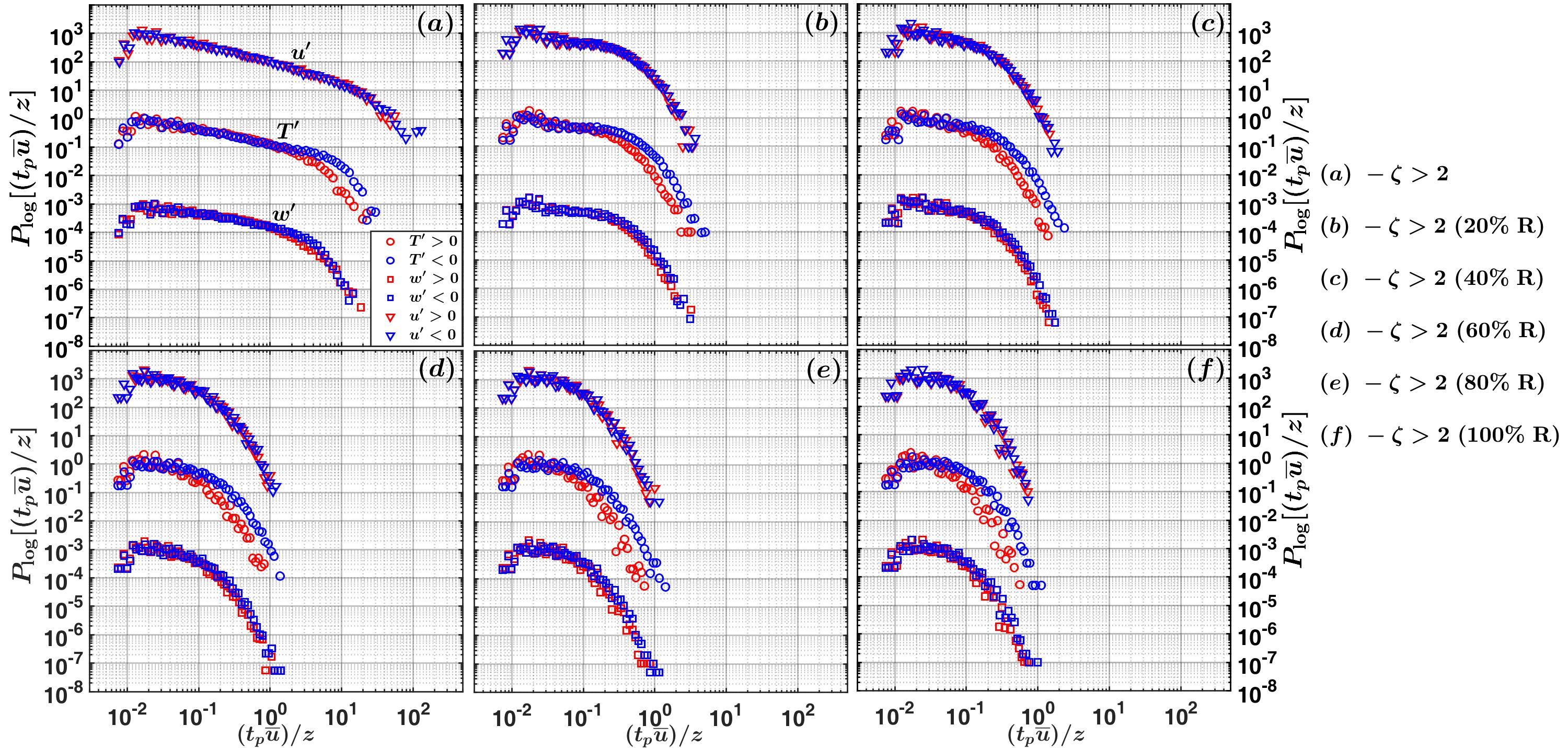}
  \caption{Same as in figure \ref{fig:4}, but for the positive and negative fluctuations in $u^{\prime}$ and $w^{\prime}$ signals also included from the highly convective stability class ($-\zeta$>2), during the temporal-randomization experiment. The panels show the signals are gradually randomized in their temporal order, beginning from the original signals to 100\% randomized signals, in the increments of 20\%.}
\label{fig:s3}
\end{figure}

\begin{figure}[h]
\centering
\vspace*{0.5in}
\hspace*{-0.75in}
\includegraphics[width=1.25\textwidth]{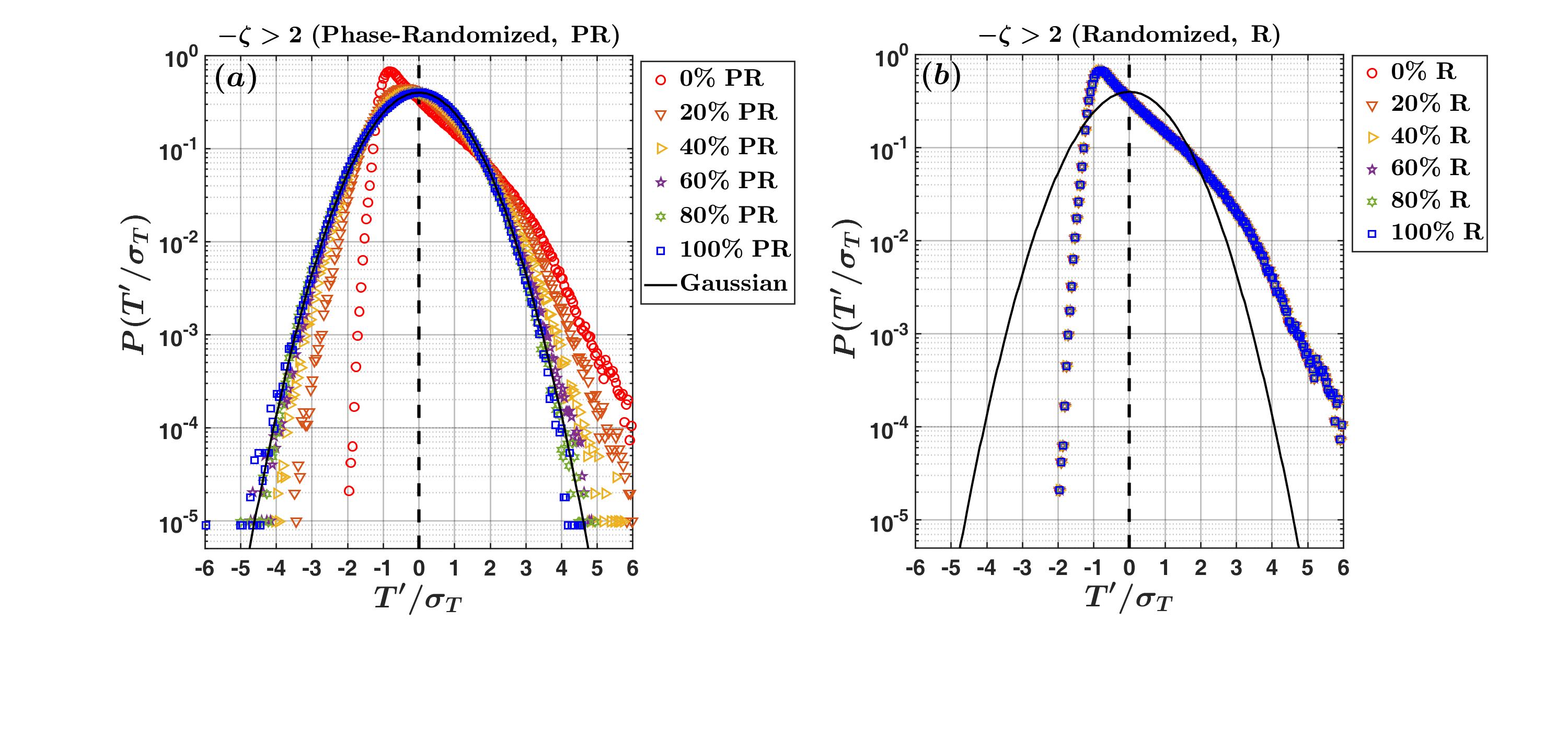}
\vspace*{-15mm}
 \caption{The PDFs of the normalized temperature fluctuations ($T^{\prime}/\sigma_{T}$) are shown from the highly convective stability class ($-\zeta>2$) for the (a) phase-randomization (PR) and (b) temporal-randomization (R) experiments. The different markers for both the panels as shown in the respective legends denote different degree of randomization, where 0\% indicates the original signal and 100\% indicates completely randomized. Note that for increasing degree of phase-randomization (see panel (a)) the original non-Gaussian temperature PDFs approach the Gaussian PDFs (shown as the thick black line), whereas the temporal-randomization preserves the PDF characteristics (see panel (b)).}
\label{fig:s4}
\end{figure}

\begin{figure}[h]
\centering
\vspace*{0.5in}
\hspace*{-2.5in}
\includegraphics[width=1.7\textwidth]{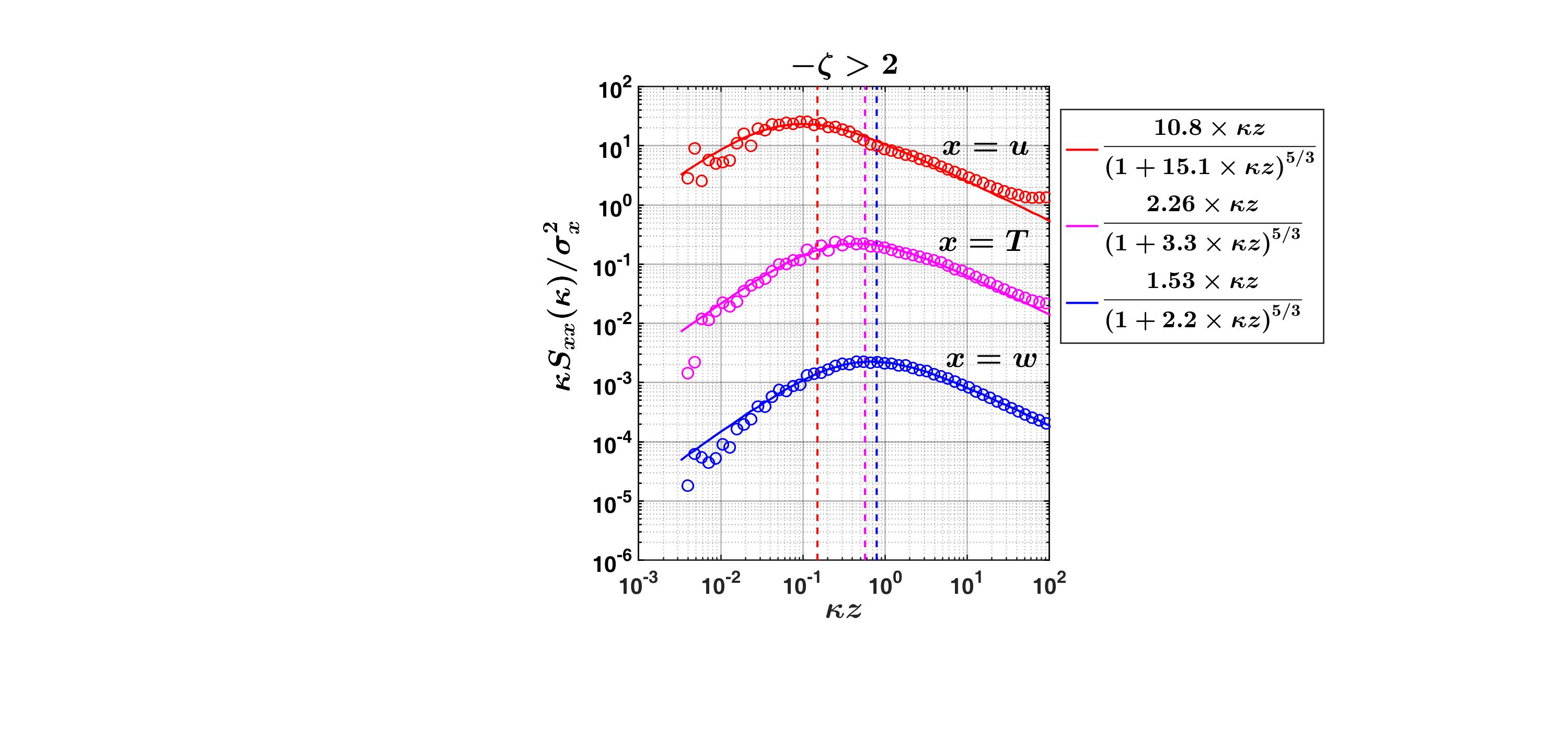}
\vspace*{-20mm}
\caption{The log-log plots of the normalized spectral energy densities with respect to the normalized streamwise wavenumbers ($\kappa z$) are shown for the $u^{\prime}$ (red circles), $T^{\prime}$ (pink circles), and $w^{\prime}$ (blue circles) signals from the highly convective stability class ($-\zeta>2$), with each being displaced vertically by the same amount for visualization purpose. Note that the spectral energy densities for each signal are normalized by their respective variances, such that the area under each curve is exactly 1. The fitted equations from the spectral model of \citet{kaimal1994atmospheric} are shown for each of the signals as indicated in the legend placed at the right-most corner. The dashed vertical lines indicate the values of $z/\Lambda_{x}$ ($x$ can be either $u$ (red), $T$ (pink), or $w$ (blue)), where $\Lambda_{x}$ is the integral length scale obtained from the exponential fit to the auto-correlation functions shown in figure \ref{fig:5}a. An excellent match is observed between the respective spectral peak positions and the values $z/\Lambda_{x}$.}
\label{fig:s5}
\end{figure}

\begin{figure}[h]
\centering
\vspace*{0.5in}
\hspace*{-0.75in}
\includegraphics[width=1.25\textwidth]{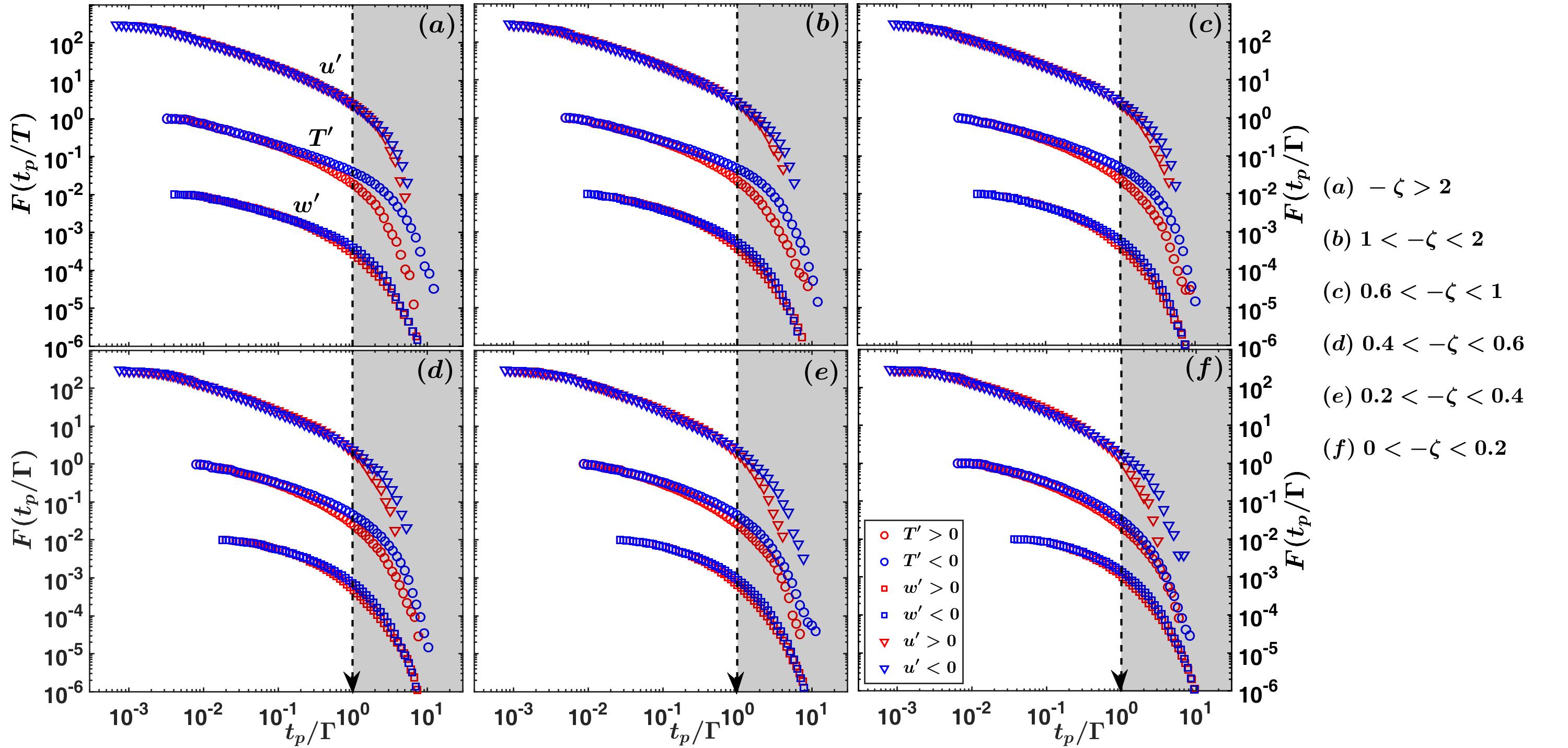}
  \caption{The cumulative distribution functions plotted against the persistence times normalized by the integral time scales ($\Gamma$) associated with $w^{\prime}$ (squares), $T^{\prime}$ (circles), and $u^{\prime}$ (inverted triangles) signals. Each panel corresponds to the six different stability classes as indicated in the right most corner and the CDFs are shifted vertically for the visualization purpose. The grey shaded regions in all the panels denote the range $t_{p}/\Gamma>$ 1.}
\label{fig:s6}
\end{figure}

\end{document}